\documentclass[11pt]{revtex4}
\usepackage{supertabular}
\usepackage{amsmath}
\usepackage{amsfonts}
\usepackage{amssymb}
\begin{document}
\title{
Discrete symmetries in the Kaluza-Klein theories}
\author{N.S. Manko\v c Bor\v stnik${}^1$ and H.B.F. Nielsen${}^2$\\
${}^1$University of Ljubljana,\\ Slovenia \\
${}^2$Niels Bohr Institute,\\
Denmark
}

\begin{abstract} 
In theories of the Kaluza-Klein kind there are spins or total angular moments  
in higher dimensions which manifest as charges in the observable $d=(3+1)$. The charge 
conjugation requirement, if following the prescription in ($3+1$), would transform any 
particle state out of the Dirac sea into the hole in the Dirac sea, which manifests as 
an anti-particle having all the spin degrees of freedom  in $d$, except $S^{03}$, the 
same as the corresponding particle state. This is in contradiction with what we observe 
for the anti-particle. In this paper we redefine the discrete symmetries so that we stay 
within the subgroups of the starting group of symmetries, while we require that the 
angular moments in higher dimensions manifest as charges in $d=(3+1)$. 
We pay attention on spaces with even $d$. 
\end{abstract}
%



\maketitle
\section{Introduction}
\label{introduction}

Since the theorem of CPT is  general under the assumption of the Lorentz 
invariance and causality~\footnote{The conservation of the product of all three 
symmetries $CPT$ is discussed in the refs.~\cite{CPTgeneral}. } it will be true in 
a world with a higher number of dimensions than the empirical 
$(3+1)$, independent of the details of the way the extra dimensional space is 
realized in such a ``Kaluza Klein theory'' as long as the assumption of the 
Lorentz invariance and causality is valid. Under these conditions the CPT symmetry 
is the symmetry of the system whatever are the extra dimensional space  details.

The concept of what the other symmetries C , P and T separately mean is in effective 
theories somewhat a matter of definition partly arranged so as to make them conserved
if possible. A theory, which would in the low energy regime explain all the 
observed phenomena, are expected, however, to have the concept of the discrete symmetries 
well understood. 

The main questions to be discussed in this article are:
\begin{itemize}
\item The definition of the discrete symmetries to be discrete symmetries 
in the higher dimensional space-time of the Kaluza Klein type, we shall denote these 
symmetries by ${\cal C}_{\cal H}$, $ {\cal P}_{\cal H}$ and ${\cal T}_{\cal H}$,   
which means that we require extension of the so far defined discrete symmetries.
\item The definition of the discrete symmetries in the $(3+1)$ dimensions  after letting 
a series or rather a group of Killing transformations to manifest the corresponding 
Noether's charges in $(3+1)$, we shall denote these symmetries by ${\cal C}_{\cal N}$, $ 
{\cal P}_{\cal N}$ and ${\cal T}_{\cal N}$, which means that we analyse the type of 
symmetries in the extra dimensional space leading to observed symmetries in $(3+1)$. 
%
\end{itemize}

There are two special examples of spaces with extra dimensions to the observed 
$(3+1)$ on which we discuss the here proposed discrete symmetries:\\ 
{\bf I.} The space of $M^{5+1}$ which breaks into $M^{3+1} \times M^2$  with $M^2$ which 
due to the zweibein compactifies in an almost $S^2$~\footnote{ We showed in the refs.~\cite{NH,NH1} 
that in such an almost compactified space the appropriately chosen spin connections 
guarantees that ($2^{\frac{d-2}{2}-1}=2$ families of) only (either) left (or right) 
handed spinors keep masslessness while being coupled with the Kaluza-Klein $U(1)$ charge 
to the corresponding gauge fields.}. 
Both, spin connections and vielbeins, have the rotational invariance around the axes 
perpendicular to the $M^2$ surface, manifesting correspondingly the $U(1)$ charge in $d=(3+1)$.\\ 
{\bf ii.} The space of $M^{13+1}$ which breaks into $M^{3+1} \times $ the rest%
~\footnote{First the manifold $M^{13+1}$ breaks into $M^{7+1}\times $ $M^{6}$,
$M^{6}$ manifesting the Kaluza-Klein charges of $SU(3)\times U(1)$, with  (eight 
families of) massless spinors, and then further to $M^{3+1}\times $ $M^{4} \times $
$M^{6}$, manifesting the symmetry of $SO(3,1)\times$ $ SU(2)\times$ $SU(2) \times U(1)$
$\times SU(3)$. Further breaks bring masses to eight families~\cite{Snmb2013Bled,NJMP,NPLB} of 
spinors. These further breaks could go similarly as it does in some theories with the 
sigma model action~\cite{BRG}. These studies are in progress.}, manifesting again 
rotational symmetries responsible for the charges in $d=(3+1)$, required by the 
{\it standard model}.  There are vielbein and spin connection fields in $d>4$ which 
manifest in $d=(3+1)$ as the corresponding gauge vector (and scalar~\cite{NPLB,NJMP}) fields 
after the compactification.

In the Kaluza-Klein kind~\footnote{With the Kaluza-Klein type of theories we mean the theories in 
which fermions carry only spins, and (may be) family quantum numbers, as internal degrees of freedom 
and interact correspondingly only through spin connections and vielbeins.} of 
theories~\cite{kk,zelenaknjiga,witten,sugra} total angular moments in higher dimensions ($d>(3+1)$) 
manifest as  charges in $(3+1)$ and the corresponding spin connections and vielbeins as the gauge 
fields~\cite{zelenaknjiga,mil}. In the low energy regime there are indeed the spin degrees of 
freedom~\footnote{The lowest energy state in any bound system is (almost always) the state with 
orbital excitation equal to zero. Like it is the $1s$ state of the hydrogen atom. A state which has 
no orbital or radial excitation when $M^{(d-1)+1}$ breaks into $M^{3+1}\times $ the rest could  
manifest the 
subgroups of the spin degrees of freedom in higher dimension as charges in $(3+1)$. As an example 
let us cite a toy model~\cite{NH,DHN}, where the rest is the infinite disc curled into an almost $S^2$. 
The lowest energy state, which appears to be massless, has the orbital angular momentum equal to zero, 
so that it is the spin in $d=(5,6)$, which manifests as the charge in the Kaluza-Klein sense. 
For all the other states, which are massive, there are subgroups of the total angular moments in higher 
dimensional space which determine the Kaluza-Klein charges in $d=(3+1)$.}
which manifest as the conserved Kaluza-Klein charges~\cite{NH,DHN}.

There are several papers~\cite{cptKK} discussing discrete symmetries in higher dimensional spaces 
in several contexts. Authors discuss mostly only the parity symmetry, some of them  
the charge  conjugation and very rare all the three symmetries. All discussions on discrete 
symmetries concern particular models. 
We are proposing the definition of the discrete symmetries for the Kaluza-Klein kind of theories 
in even dimensional spaces~\footnote{We demonstrate in the refs.~\cite{NH1} that the masslessness 
of fermions can be guaranteed only in even dimensional spaces.}. 
This definition leads after compactification of 
space-time into the (so far) observed $(3+1)$-dimensional space to the measured properties of 
particles and anti-particles.

{\em Extending the prescription of the discrete symmetries from $d= (3+1)$ to any $d$ }
(Eq.~(\ref{calCH}, \ref{calTPH}), sect. \ref{dh}), {\em the anti-particle to a chosen particle would 
have in the second quantized theory all the components of spin, or total 
angular momentum }(except the $S^{03}$ component which is involved in the boost and does not contribute 
to the spin component; in quantum mechanics time is a parameter), {\em the same as the starting particle}, 
which means that {\em it would have all the charges the same as the corresponding particle}. This would 
be in contradiction with what we observe, namely that the anti-particle to a chosen particle has opposite 
charges. 

In this paper, sect. \ref{dn}, we modify the $d$-dimensional discrete symmetries, for example the charge 
conjugation operator 
${\cal C}_{{\cal H}}$ (Eq.~(\ref{calCH})) as it would follow from the $(3+1)$ case  by analogy, 
so that they work effectively in the $(3+1)$ dimensional  theory. As we shall see below, the connection 
between the effective three dimensional ones~(Eqs.~(\ref{CNsq},  \ref{CNsqshorter0})), 
${\underline {\bf \mathbb{C}}}_{{\bf {\cal N}}}$, 
${\cal T}_{{\cal N}}$ and ${\cal P}^{(d-1)}_{{\cal N}}$, and the $d$-dimensional 
ones~(Eq.~(\ref{CHsq}, \ref{CsqCf})), ${\underline {\bf \mathbb{C}}}_{{\bf {\cal H}}}$, 
 ${\cal T}_{{\cal H}}$ and ${\cal P}^{(d-1)}_{{\cal H}}$, is a
multiplication with products of representatives of the Lorentz group corresponding to reflections 
and a parity operator 
in higher dimensions. Our notation is that we put index ${\cal H}$ on discrete symmetries 
${\cal P}^{(d-1)}_{{\cal H}}$, ${\cal T}_{{\cal H}}$, and ${\cal C}_{{\cal H}}$
for the whole space, i.e. $d$ dimensions, while we use ${\cal N}$ for the effective discrete 
symmetries in only our $(3+1)$ dimensions.
We define three kinds of the charge conjugation operator: ${\cal C}_{({\cal H,N})}$, 
$\mathbb{C}_{({\cal H,N})}$, and ${\underline {\bf \mathbb{C}}}_{({\bf {\cal H,N}})}$. The first one 
operates on the single particle state, put on the top of the Dirac sea, transforming the positive 
energy state  into the corresponding negative energy state~(Eqs.(\ref{calCH}), (\ref{CNsq})). 
The second one does the job of the first one emptying~\cite{NJMP,TDNBled2013} (Eqs.(\ref{makingantip}), 
(\ref{emptcalCH}), (\ref{CNsq}), (\ref{CNsqshorter0})) in addition the negative energy state,  
creating correspondingly a hole, which manifests as a positive 
energy anti-particle state, put on the top of the Dirac sea. (The corresponding single anti-particle state
must also solve the equations of motion as the starting particle state does, although we must 
understand it as a hole in the Dirac sea in the context of the Fock space). The third 
one~Eqs.((\ref{makingantip}), (\ref{CNsq}))  is the operator, operating on the second quantized 
state~(Eq.(\ref{CsqCf})).

Discrete symmetries presented in this paper commute with the family quantum numbers - the family groups 
defining the equivalent representations with respect to the spin and correspondingly to all the charge 
groups have no influence on the here presented discrete symmetries~\footnote{This paper is initiated 
by the theory, proposed by one of us 
(S.N.M.B)~\cite{norma,Portoroz03,pikanorma,Snmb2013Bled,NJMP,NPLB}, and  named the 
{\it spin-charge-family} theory. This theory,  
which is offering the mechanism for generating families, predicts consequently the number of 
observable families at low  energies. It also predicts several scalar fields which at low energies 
manifest as the Higgs and Yukawa couplings of the {\it standard model}. The spin in higher 
dimensions manifests as the observed charges in $d=(3+1)$, as in all the Kaluza-Klein kind of theories.
The generators of the groups, determining families in this theory, commute with the total angular 
momentum in all  dimensions. 
Both authors have published together several papers, proving that in non-compact spaces the break 
of the starting symmetry in $d>4$ might allow massless fermions after the break~\cite{DHN} for all the 
Kaluza-Klein theories.}.

Although we illustrate our proposed discrete symmetries in two special cases (sects.~\ref{freespinH}, 
\ref{freespinN}, \ref{interspin}, \ref{understanding}), in which fermions,    
 sect.~\ref{interspin},  interact in the  Kaluza-Klein way with the vielbein and spin connection fields, 
 the proposed redefinition of the discrete symmetries, marked by index ${\cal N}$, is  expected to be 
 quite general, offering experimentally 
  observed properties of anti-particles in $d=(3+1)$ for the Kaluza-Klein kind of theories, 
  helping also to define the discrete symmetries in $d=(3+1)$ in other cases with higher dimensional spaces.
  
  We allow, in general,  curling up extra dimensions by various bosonic background fields (metric tensors, 
  magnetic fields, .. in extra dimensions) as far as the equations of motion  determining properties of fermions 
  in extra dimensions keep the proposed discrete symmetries conserved.
  
  We assumed that there are  a few fixed points symmetries and particular rotational symmetries 
  around these fixed points in higher than $d=(3+1)$  and that there are a series of Cartan subalgebra 
  symmetries around  fixed points.  Various subgroups of the rotations around (a) fixed point(s)
  are the "Killing forms" manifesting charges in the $(3+1)$ effective theory. (The example of compactified 
  two extra dimensions on an almost $S^2$-sphere with a Killing form transformation being a rotation 
  of the sphere illustrates that typically there shall be two fixed points. But if we had for instance an infinite 
  extra dimensional space, only one fixed point is also possible.)
  
  Having such one or more fixed points attached to the "Killing forms" of the charges makes it very attractive 
  and natural to  assume parity symmetry under point inversions in the fixed point(s)
  (the parity operation should at the same time be inversion in both fixed points, if, say, there are two).
  Combining such a suggestively imposed parity inversion in the extra dimensions with the 
  parity operation in $(3+1)$ would lead to parity operation ${\cal P}^{(d-1)}_{{\cal H}}$ (Eq.~(\ref{calTPH})) 
  in all the $(d-1)$ spatial dimensions. 
  
  {\em Our effective parity}, ${\cal{P}}^{d-1}_{\cal{N}}$, Eqs.~(\ref{CNsq}, \ref{CNsqshorter0}), {\em proposal}
  does, however,  {\em not contain any 
  transformation of the extra dimensional coordinates} and just  {\em got the contribution }of the $\gamma^{a}$ 
  matrices {\em adjusted so that the extra dimensional gamma matrices } $\gamma^5$, $\gamma^6$, ..., 
  $\gamma^{d-1}$,$\gamma^d$ {\em commute with }
  ${\cal{P}}^{d-1}_{\cal{N}}$. {\em This means that this operation is quite insensitive to the extra dimensions}
  in such a way that it is not important if the extra dimensional space obeys any parity like symmetry.  
  
   We pay attention on spaces with even $d$~\footnote{We do not pay attention on renormalizability of 
 the theory in this paper.}.

 We do not discuss the way  how does an (almost) compactification happen in our here discussed 
 two particular cases. 
 In the ref.~\cite{DHN} we propose   vielbein and spin connection fields which are responsible for the 
 compactification of an infinite surface 
 into an almost $S^{2}$, but do not tell what  (fermion condensates)  causes the 
 appearance of these  gauge fields. These studies are for the two cases, presented in this paper, 
 under consideration. 
 There are, however,  several proposals in the literature which suggest the compactification scheme and 
 discuss it~\cite{CEZ2013}. We are not yet able to comment them from the point of view of our two  
 discussed cases.
 
Our new discrete symmetries are demonstrated in sect.~\ref{dn}, in which spins or total angular 
moments in higher dimensions manifest   charges of massless and massive spinors in $d=(3+1)$, 
by showing how  do the example wave functions and quite general Lagrange density transform under 
the ${\cal C}_{({\cal H,N})}$, ${\cal P}^{(d-1)}_{({\cal H,N})}$,  and ${\cal T}_{({\cal H,N})}$ 
discrete symmetries. 
These two particular cases concern  fermions
, the charges of 
which originate in $d>4$, in: i.) $d=(5+1)$, when $SO(5,1)$ breaks into $SO(3,1) \times U(1)$~\cite{DHN}, 
with $U(1)$ manifesting as the Kaluza-Klein charge in $d=(3+1)$. ii.) $d=(13+1)$, the symmetry of 
which breaks into $SO(3,1)\times SU(3) \times SU(2)$ $\times U(1)$, 
while the subgroups determine charges of  fermions, manifesting before the electroweak break left 
handed weak charged and right handed weak chargeless massless quarks and 
leptons~\cite{norma,Portoroz03,pikanorma,gmdn,gn,BledAN2010,NJMP,NPLB} of the {\em standard model}. 
In these two demonstrations 
the technique~\cite{norma93,hn0203} is used to treat spinor degrees of freedom, which is very convenient 
for this purpose, since it is transparent and simple. 

We discuss the generality of our effective proposal for discrete symmetries in section~\ref{understanding}, 
in subsection~\ref{comp5+1} of which we discuss our two special cases, commenting also possible way of 
compactifying the higher dimensional space.

We shall use the concept of the Dirac sea second quantized picture, which is equivalent to the formal 
ordinary second quantization, because it offers, in our opinion, a nice physical understanding.  


We do not study in this paper the break of the $CP$ and correspondingly of the $T$ symmetry.

\section{Discrete symmetries 
in d-dimensions following the definitions in $d=(3+1)$}
\label{dh}

We start with the  definition of the discrete symmetries  as they follow from the prescription
in  $d=(3+1)$. We treat particles which carry in $d$ dimensions only spin, no charges. They also carry 
the family quantum numbers, which, however, commute with the discrete family operators.  

We first treat free spinors.
We define the ${\cal C}_{{\cal H}}$ operator  to be distinguished from  the $\mathbb{C}_{{\cal H}}$ 
operator. The first transforms any single particle state $\Psi^{pos}_{p}$, index ${}_{p}$ denotes the 
fermion state, which solves the Weyl equation for a 
free massless spinor with a positive energy and it is in the second quantized theory understood as the state 
above the Dirac sea, into the charge conjugate one with the negative energy $\Psi^{neg}_{p}$ and 
correspondingly belonging to a state in the Dirac sea
\begin{eqnarray}
\label{calCH}
{\cal C}_{{\cal H}}= \prod_{\gamma^a \in \Im} \gamma^a \,\, K\,.
\end{eqnarray}
The product of the imaginary $\gamma^a$ operators is meant in the ascending order. We make a choice 
of $\gamma^0, \gamma^1$ real, $\gamma^2$ imaginary, $\gamma^3$ real, $\gamma^5$ imaginary, $\gamma^6$ real, 
and alternating real and imaginary ones we end up in even dimensional spaces with real $\gamma^d$. 
$K$ makes complex conjugation, transforming $i$ into $-i$.

We define $\bf{\mathbb{C}_{{\cal H}}}$ as the operator, which   {\it emptyies the negative energy  state 
in the Dirac sea following from the starting positive energy state, and  creates an anti-particle with 
the positive energy and all the properties of the starting single particle state above the Dirac sea -  
that is with the same  $d$-momentum and all the spin degrees of freedom the same, except the $S^{03}$ 
value}, as the starting single particle state. The operator $
S^{03}$ is involved in the boost (contributing in $d=(3+1)$, together with the spin, to handedness) and does not 
determine the (ordinary) spin. Accordingly  we do not have to keep the $S^{03}$ value a priori unchanged under 
the charge conjugation. Had we instead considered $CP$ we would also have kept $S^{03}$. 

Let  ${\underline {\bf {\Huge \Psi}}}^{\dagger}_{p}[\Psi^{pos}_{p}]$ be the creation  operator  creating  
a fermion in the state $\Psi^{pos}_{p}$ (which is a function of $\vec{x}$) and let 
${\mathbf{\Psi}}_{p}(\vec{x})$
be the second quantized field creating a fermion at position $\vec{x}$. Then 
\begin{eqnarray}
\label{CHsq}
{\underline {\bf {\Huge \Psi}}}^{\dagger}_{p}[\Psi^{pos}_{p}] &=& \int \, {\mathbf{\Psi}}^{\dagger}_{p}(\vec{x})\,
\Psi^{pos}_{p}(\vec{x}) d^{d-1} x 
\end{eqnarray}
or on a vacuum where it describes a single particle in the state $\Psi^{pos}$
\begin{eqnarray}
\{ {\underline {\bf {\Huge  \Psi}}}^{\dagger}_{p}[\Psi^{pos}_{p}] &=& \int \, 
{\mathbf{\Psi}}^{\dagger}_{p}(\vec{x})\, 
\Psi^{pos}_{p}(\vec{x}) d^{d-1} x \,\} 
\,|vac> \nonumber
\end{eqnarray}
so that the anti-particle  state becomes
\begin{eqnarray}
\{ {\underline {\bf \mathbb{C}}}_{{\bf {\cal H}}}\, 
{\underline {\bf {\Huge \Psi}}}^{\dagger}_{p}[\Psi^{pos}_{p}] &=&  
\int \, {\mathbf{\Psi}}_{p}(\vec{x})\, 
({\cal C}_{{\cal H}} \,\Psi^{pos}_{p} (\vec{x})) d^{d-1} x \} \, |vac> \,.\nonumber
\end{eqnarray}
We also can derive the relation
\begin{eqnarray}
\label{CsqCf}
 {\underline {\bf \mathbb{C}}}_{{\bf {\cal H}}}\,  {\mathbf{\Psi}}(\vec{x})\,
({\underline {\bf \mathbb{C}}}_{{\bf {\cal H}}})^{-1}
 &=& {\cal C}_{{\cal H} formal}  \, 
 {\mathbf{\Psi}}(\vec{x})=({\cal C }_{{\cal H}}\,K)_{ formal}  \, {\mathbf{\Psi}}^{\dagger}(\vec{x})\,.
\end{eqnarray}
This formal operation ${\cal C}_{{\cal H} formal}$  means the action on the second quantized field  
${\mathbf{\Psi}}$ as if it were 
a function of $\vec{x}$ and a column in gamma matrix space, and that the complex conjugation 
is replaced by the Hermitian 
conjugation (${}^{\dagger}$) on the second quantized operator~\footnote{This simply means that, 
for example, we can use 
Hermitian conjugate equations of motion for 
$({\cal C }_{{\cal H}}\,K)_{ formal} \,{\mathbf{\Psi}}(\vec{x})$ and then 
check the ${\cal C }_{{\cal H}}$ without  the complex conjugation: $({\cal C }_{{\cal H}}\,K)_{ formal}$. }. 

Let us define the operator "emptying"~\cite{Snmb2013Bled,TDNBled2013} the Dirac sea, so that operation of 
"emptying" after 
the charge conjugation ${\cal C }_{{\cal H}}$ (which transforms the state put on the top of the Dirac sea 
into the corresponding negative energy state) creates the anti-particle state to the starting 
particle state, both  put on the top of the Dirac sea and both solving the Weyl equation for a free massless 
fermions
\begin{eqnarray}
\label{empt}
"emptying"&=& \prod_{\Re \gamma^a}\, \gamma^a \,K =(-)^{\frac{d}{2}+1} \prod_{\Im \gamma^a}\gamma^a \,
\Gamma^{(d)} K\,, 
\end{eqnarray}
although we must keep in mind that indeed the anti-particle state is a hole in the Dirac sea from the 
Fock space point of view. The operator "emptying" is bringing the single particle operator 
${\cal C }_{{\cal H}}$ into the operator on the Fock space.
Then the anti-particle  state creation operator - 
${\underline {\bf {\Huge \Psi}}}^{\dagger}_{a}[\Psi^{pos}_{p}]$ - to the corresponding  particle state 
creation operator - can be obtained also as follows
\begin{eqnarray}
\label{makingantip}
{\underline {\bf {\Huge \Psi}}}^{\dagger}_{a}[\Psi^{pos}_{p}]\, |vac>  &=& 
{\underline {\bf \mathbb{C}}}_{{{\bf \cal H}}}\, 
{\underline {\bf {\Huge \Psi}}}^{\dagger}_{p}[\Psi^{pos}_{p}]\, |vac>  =  
\int \, {\mathbf{\Psi}}^{\dagger}_{a}(\vec{x})\, 
({\bf \mathbb{C}}_{\cal H}\,\Psi^{pos}_{p} (\vec{x})) \,d^{d-1} x  \, \,|vac> \,,\nonumber\\
{\bf \mathbb{C}}_{\cal H} &=& "emptying"\,\cdot\, {\cal C}_{{\cal H}}  \,.
\end{eqnarray}
The operator ${\bf \mathbb{C}}_{\cal H} = "emptying" \,\cdot\, {\cal C}_{{\cal H}}$ operating on 
$\Psi^{pos}_{p} (\vec{x})$ transforms the positive energy spinor state (which solves the Weyl equation 
for a massless free spinor) put on the top of the Dirac sea into the positive energy anti-spinor 
state, which again solves the Weyl equation for a massless free anti-spinor put on the top of 
the Dirac sea. Let us point out that the operator $"emptying" $ transforms the single particle operator
$ {\cal C}_{{\cal H}}$ into the operator operating in the Fock space. 

The operator "emptying"  operates meaningfully in all known cases when the higher dimensions manifest 
charges or masses or both in $d= (3+1)$ space.

We define the time reversal operator ${\cal T}_{{\cal H}}$ and the  parity operator 
${\cal P}^{(d-1)}_{{\cal H}}$ as follows 
\begin{eqnarray}
\label{calTPH}
{\cal T}_{{\cal H}}&=& \gamma^0 \prod_{\gamma^a \in \Re} \gamma^a \,\, K\, I_{x^0}\,,\nonumber\\
{\cal P}^{(d-1)}_{{\cal H}} &=& \gamma^0\,I_{\vec{x}}\,,\nonumber\\
I_{x} x^a &=&- x^a\,, \quad I_{x^0} x^a = (-x^0,\vec{x})\,, \quad I_{\vec{x}} \vec{x} = -\vec{x}\,, \nonumber\\
I_{\vec{x}_{3}} x^a &=& (x^0, -x^1,-x^2,-x^3,x^5, x^6,\dots, x^d)\,.
\end{eqnarray}
Again the product $\prod \, \gamma^a$ is meant in the ascending order in $\gamma^a$.

Let us calculate now the product of ${\cal C}_{{\cal H}}$ ${\cal P}^{(d-1)}_{{\cal H}}$ ${\cal T}_{{\cal H}}$
and ${\bf \mathbb{C}}_{\cal H}$ $ {\cal P}^{(d-1)}_{\cal H} $ ${\cal T}_{\cal H}$
\begin{eqnarray}
\label{calCTPH}
{\cal C}_{\cal H} {\cal P}^{(d-1)}_{\cal H} {\cal T}_{\cal H}&\propto & \Gamma^{(d)}\;I_{x}\,,\nonumber\\
{\bf \mathbb{C}}_{\cal H} {\cal P}^{(d-1)}_{\cal H} {\cal T}_{\cal H}&=&"emptying"\,\cdot \,
{\cal C}_{\cal H} {\cal P}_{\cal H} {\cal T}_{\cal H}\propto \prod_{\gamma^a \in \Im } \gamma^a \,\,I_{x} \,K\,,
\end{eqnarray}
with
\begin{eqnarray}
\label{emptcalCH}
{\bf \mathbb{C}}_{\cal H}&=&  \prod_{\Re \gamma^a}\, \gamma^a \,K \,{\cal C}_{\cal H}\,\propto  \Gamma^{(d)}\,.
\end{eqnarray}
$\propto$ stays for up to a phase. It follows
\begin{eqnarray}
\label{emptmathCPTH}
{\underline {\bf \mathbb{C}}}_{{{\bf \cal H}}} {\cal P}^{(d-1)}_{\cal H} {\cal T}_{\cal H}\,
{\underline {\bf {\Huge \Psi}}}^{\dagger}_{p}[\Psi^{pos}_{p}]\,
({\underline {\bf \mathbb{C}}}_{{{\bf \cal H}}}{\cal P}^{(d-1)}_{\cal H} {\cal T}_{\cal H})^{-1}
&=&{\underline {\bf {\Huge \Psi}}}^{\dagger}_{a}
[{\bf \mathbb{C}}_{\cal H} {\cal P}^{(d-1)}_{\cal H} {\cal T}_{\cal H}\,\Psi^{pos}_{p}]\,.
\end{eqnarray}
$\Gamma^{(d)}$ is defined in Eq.~(\ref{Gamma})

\subsection{Free spinors case}
\label{freespinH}

To demonstrate what do the discrete symmetry operators of Eqs.~(\ref{calCH}, \ref{calTPH}, 
\ref{emptcalCH}) do on the spinor states let us first look for the solutions of the Weyl equation 
for a free spinor in $d=(d-1)+1$ for $d$ even,  
\begin{eqnarray}
\label{weyl}
\gamma^a p_a \,\psi=0\,,
\end{eqnarray}
and show the application of the above defined discrete symmetries on the solutions for two particular 
cases: {\bf i}.  $d=(5+1)$, the properties of which we study in several papers~\cite{NH,DHN}, and  {\bf ii.} 
$d=(13+1)$,  which one of the authors of this paper uses  in her {\it spin-charge-family} 
theory~\cite{norma,Portoroz03,pikanorma,gmdn,gn,BledAN2010,NJMP,NPLB}, since it manifests in $d=(3+1)$ in 
the low energy regime 
the family members  (explaining correspondingly the appearance of families) with  the family members   
assumed by the {\it standard model} (extended with the right handed neutrino).  
Let us recognize that the operator of handedness, expressed in terms of the Cartan subalgebra members,
is as follows
\begin{eqnarray}
\label{Gamma}
\Gamma^{((d-1)+1)} = (-2i)^{\frac{d}{2}}\, S^{03} S^{12} S^{56}\dots S^{(d-1)d}\,.
\end{eqnarray}

For the choice of the coordinate system so that $d$-momentum manifests 
 $p^a= (p^0,0,0,p^3,0 \dots 0)$ the Weyl equation simplifies to
\begin{eqnarray}
\label{weylsimpl}
(-2i S^{03} p^0= p^3)\psi \,.
\end{eqnarray}
We shall make use of this choice. 
Solutions in the coordinate representation are plane waves: $e^{-ip^a x_a}$. In this part
${\cal T}_{{\cal H}}$ and ${\cal P}_{{\cal H}}$ manifest as follows
\begin{eqnarray}
\label{planewaveTP}
{\cal T}_{{\cal H}} (\cdots) e^{-ip^0 x^0 + ip^3 x^3} = (\cdots) e^{-ip^0 x^0 - ip^3 x^3}\,,\quad\quad
{\cal P}_{{\cal H}} (\cdots) e^{-ip^0 x^0 + ip^3 x^3} = (\cdots) e^{-ip^0 x^0 - ip^3 x^3}\,,
\end{eqnarray}
since in the momentum representation only $p^a$ is a vector, while $x^a$ is just a parameter 
(and opposite in the coordinate representation). (With ${\cal T}_{{\cal H}}$ transformed wave 
function develops the usual Schroedinger way for $x^0$ is replaced by $-x^0$.)

{\bf $d=(5+1)$ case}:\\
Let us now demonstrate the application of the discrete operators 
${\underline {\bf \mathbb{C}}}_{{{\bf \cal H}}}$, 
${\cal T}_{{\cal H}}$ and ${\cal P}_{{\cal H}}$ on one Weyl representation from 
Table~\ref{Table I.}, which represents the positive and negative energy solutions of the Weyl 
equation~(\ref{weylsimpl}) in $d=(5+1)$. Here and in what follows we do not pay attention 
on the normalization factor of the single particle states.
Let us make a choice of the positive energy state $\psi^{pos}_{1}=$
$\stackrel{03}{(+i)} \stackrel{12}{(+)}\stackrel{56}{(+)} e^{-ip^0 x^0 +ip^3 x^3}$, for example.
We use the technique of the refs.~\cite{hn0203,norma93}. 
A short overview can be found in the Appendix. The reader
is kindly asked to look for more detailed explanation in~\cite{hn0203}.
It follows for $p^0 = |p^0|$ and $p^3 = |p^3|$
\begin{eqnarray}
\label{calC1}
{\cal C}_{{\cal H}}\psi^{pos}_{1}\rightarrow \stackrel{03}{(+i)} 
\stackrel{12}{[-]}\stackrel{56}{[-]} e^{ip^0 x^0 -ip^3 x^3}=
\psi^{neg}_{2}\,.
\end{eqnarray}
This state is the solution of the Weyl equation for the negative energy state. But the hole 
of this state in the Dirac sea makes a positive energy state (above the Dirac sea) with the {\em properties} of 
the starting state, but {\em it is an anti-particle} state: $\Psi^{ pos}_{a1}=$
$\stackrel{03}{(+i)} \stackrel{12}{(+)} \stackrel{56}{(+)} e^{-ip^0 x^0 +ip^3 x^3}$, 
defined~\footnote{If one would like a more detailed meaning of $\Psi^{pos}_{a}$ one can imagine the 
second quantization 
of the whole theory using anti-particles instead of particles in the  theory and so obtaining the 
original particles as holes. In such a theory an anti-particle state corresponding to $\Psi^{pos}_{a}$ 
would be ${\protect\underline{\bf {\Huge \Psi}}}^{\dagger} [\Psi^{pos}] |anti vac>$, 
therefore ${\protect\underline{\bf {\Huge \Psi}}}^{\dagger} [\Psi^{neg}] |anti vac>$
$\rightarrow $
${\protect\underline{\bf {\Huge \Psi}}}^{\dagger} [\Psi^{pos}] |anti vac>$.} 
on the Dirac sea with the hole belonging to the negative energy single-particle state $\psi^{neg}_{2}$. 
Namely, ${\underline {\bf \mathbb{C}}}_{{{\bf \cal H}}}\, {\underline {\bf {\Huge \Psi}}}[\Psi^{pos}_{p}]\, 
{\underline {\bf \mathbb{C}}}_{{{\bf \cal H}}}^{-1}$, 
when applied on the 
vacuum state, represents an anti-particle. 

This anti-particle state  is correspondingly the  solution of the same Weyl equation, and it belongs 
to the same representation as the starting state (and  ${\underline {\bf \mathbb{C}}}_{{{\bf \cal H}}}$ 
is obviously a good symmetry in this $d=\,2\,(\mod 4)$ space). 
The operator $\mathbb{C}_{ \cal H}$ from Eq.~(\ref{emptcalCH}), 
applied on the state $\psi^{pos}_{p1}$, gives the same result: 
 $\psi^{pos}_{a1} $, which belong to the same representation of the Weyl equation as the starting state. 
But this state has the $S^{56}$ spin, which should 
represent in $d=(3+1)$ the charge of the anti-particle,  the same as the starting state. This is not in 
agreement with what we observe.

Since both
${\cal T}_{{\cal H}}$ (${\cal T}_{{\cal H}}\psi^{pos}_{1}$ $= \stackrel{03}{[-i]} 
\stackrel{12}{[-]}\stackrel{56}{[-]} e^{-ip^0 x^0 -ip^3 x^3}$)
and ${\cal P}_{{\cal H}}$ (${\cal P}_{{\cal H}}\psi^{pos}_{1}$ $= \stackrel{03}{[-i]} 
\stackrel{12}{(+)}\stackrel{56}{(+)} e^{-ip^0 x^0 -ip^3 x^3}$)  
are defined with an odd number 
of $\gamma^a$  operators, none of them are the symmetry (the conserved operators) within one Weyl 
representation, since both transform  correspondingly the starting state into a state of another 
Weyl representation. (This is true for all the spaces with $d=\,2\,(\mod 4)$, while in the spaces
with $d=\,0\,(\mod 4)$ the operator  ${\cal T}_{{\cal H}}$ has an even product of $\gamma^a$, while 
${\cal C}_{{\cal H}}$ contains an odd number of $\gamma^a$.)

The product of ${\cal T}_{{\cal H}}$ and ${\cal P}^{(d-1)}_{{\cal H}}$ is again a good symmetry, transforming 
the starting state, say $\psi^{pos}_{1}$, into a positive 
energy state of the same Weyl representation, 
%
${\cal T}_{{\cal H}}\,{\cal P}^{(d-1)}_{{\cal H}} \psi^{pos}_{1}= \stackrel{03}{(+i)} 
\stackrel{12}{[-]}\stackrel{56}{[-]} e^{-ip^0 x^0 + ip^3 x^3}=
\psi^{pos}_{2}\,$,
and solving the  Weyl equation.

Also the product of all three discrete symmetries is correspondingly a good symmetry as well, transforming 
the starting state (put on the top of the Dirac sea) into the positive energy anti-particle state, 
%
${\underline {\bf \mathbb{C}}}_{{{\bf \cal H}}}$
$ {\cal T}_{{\cal H}}\,{\cal P}^{(d-1)}_{{\cal H}}\, 
{\underline {\bf {\Huge \Psi}}}^{\dagger}[\Psi^{pos}_1]
({\underline {\bf \mathbb{C}}}_{{{\bf \cal H}}}$ ${\cal T}_{{\cal H}}\,{\cal P}^{(d-1)}_{{\cal H}})^{-1}\,$ 
$= {\underline {\bf {\Huge \Psi}}}^{\dagger}_{a}[\mathbb{C}_{{\cal H}}
{\cal T}_{{\cal H}}\,{\cal P}^{(d-1)}_{{\cal H}}\Psi^{pos}_{1}]
\rightarrow {\underline {\bf {\Huge \Psi}}}^{\dagger}_{a}[\Psi^{pos}_{2}]\,$,
%
which is the hole in the state $\psi^{neg}_{1}$ in the  Dirac sea.

{\bf $d=(13+1)$ case}:\\
Let us now look at 
$d=(13+1)$ case, the positive energy states of which are presented in Table~\ref{Table II.}. 
Following the procedure used in the previous case of $d=(5+1)$, the operator ${\cal C}_{{\cal H}}$
transforms, let say the first state in Table~\ref{Table II.}, which represents due to its quantum numbers
the right handed  (with respect to $d=(3+1)$) $u$-quark  with spin up, weak chargeless, carrying the 
colour charge ($\frac{1}{2}, \frac{1}{(2\sqrt{3})}$), the third component of the second $SU(2)_{II}$ 
charge $\frac{1}{2}$, the hyper charge $\frac{2}{3}$ and the electromagnetic charge $\frac{2}{3}$, while it carries the 
momentum $p^a=(p^0, 0, 0, p^3, 0, \dots,0)$, as follows
\begin{eqnarray}
\label{calC13}
{\cal C}_{{\cal H}} u_{1 R}\rightarrow \stackrel{03}{(+i)} \stackrel{12}{[-]}| \stackrel{56}{[-]} 
\stackrel{78}{[-]} ||\stackrel{9\,10}{[-]}\stackrel{11\,12}{[+]} \stackrel{13\,14}{[+]}\,
e^{ip^0 x^0 -ip^3 x^3}\,.
\end{eqnarray}
This state solves the Weyl equation for the negative energy and inverse momentum, carrying all 
the eigenvalues of the Cartan subalgebra operators ($S^{12}, S^{5 6}$, $S^{7 8}$, $S^{9\,10}$, 
$S^{11\,12}$, $S^{13\,14}$), except $S^{03}$, of the opposite values than the starting state 
(this negative energy state is a part 
of the starting Weyl representation, not presented in Table~\ref{Table II.}, but the reader can 
find this state in the ref.~\cite{Portoroz03}). 
The second quantized charge conjugation operator ${\underline {\bf \mathbb{C}}}_{{{\bf \cal H}}}$
 empties ${\cal C}_{{\cal H}} u_{1 R}$  in the 
Dirac sea, creating the anti-particle state to the starting state  with all the 
quantum numbers of the starting state, obviously in contradiction with the observations, that the 
anti-particle state has the same momentum in $d=(3+1)$ but  opposite 
charges than the starting state.

We conclude that {\it the second quantized anti-particle state} (the hole in the Dirac sea) {\it manifests 
correspondingly all the quantum numbers of the starting state, but it is the anti-particle}. 
Requiring that the eigenvalues of the Cartan subalgebra members 
in $d\ge5$ represent  charges in $d=(3+1)$, the {\it charges should have opposite values}, which the definition
of the discrete symmetries operators in Eqs.~(\ref{calCH}, \ref{calTPH}) does not offer. 
The charge conjugation operation is a good symmetry in any $d=\,2\,(\mod 4)$ from the point of view
that in any of spaces with $d=\, 2(\mod 4)\,$ $\,{\cal C}_{{\cal H}} \, \psi^{pos}_{i}$ defines the 
state within the same Weyl representation due to the fact that it is defined as the product of an 
even number of imaginary operators $\gamma^a$.
The product of the time reversal and the parity operation is in the space with $d=\,2\,(\mod 4)$ 
again a good symmetry, which means that it transforms the starting state  of a chosen Weyl representation 
into the state belonging to the same Weyl representation, with the same $d$-momentum as the 
starting state.

\subsubsection{Solutions of the Weyl equations in $d=(5+1)$}
\label{5+1}

There are $2^{\frac{d}{2}-1}= 4$ basic spinor states of one family representation in $d=(5+1)$~\footnote{
There are for $d=6$ in the {\em spin-charge-family} proposal  $2^{\frac{d}{2}-1}= 4$ families 
of spinors.}. Since the operators of 
Eqs.~(\ref{calCH}, \ref{calTPH}) do not distinguish among the families,  
all the families behave equivalently with respect to these discrete symmetry operators. 
One of the family representation, with four basic spinor states, is in the technique~\cite{hn0203}, 
described in terms of nilpotents $\stackrel{ab}{(k)}$ and projectors $\stackrel{ab}{[k]}$ 
(see Appendix~\ref{technique}),  as follows 
\begin{eqnarray}
\Psi_{1} &=&  \stackrel{03}{(+i)} \stackrel{12}{(+)} \stackrel{56}{(+)}|vac>_{fam},\nonumber\\
\Psi_{2} &=&\stackrel{03}{(+i)} \stackrel{12}{[-]} \stackrel{56}{[-]}  |vac>_{fam},\nonumber\\
\Psi_{3} &=&  \stackrel{03}{[-i]} \stackrel{12}{[-]} \stackrel{56}{(+)} |vac>_{fam},\nonumber\\
\Psi_{4} &=& \stackrel{03}{[-i]} \stackrel{12}{(+)}\stackrel{56}{[-]}|vac>_{fam}\,,
\label{weylrep}
\end{eqnarray}
where $|vac>_{fam}$ is defined so that there are  $2^{\frac{d}{2}-1}$ family members (this is, however, not 
a second quantized vacuum). 
All the basic states are eigenstates of the Cartan subalgebra (of the Lorentz transformation Lie algebra),  
for which we take: $S^{03}, S^{12}, S^{56}$, with the eigenvalues, which can be read from 
Eq.~(\ref{weylrep}) if taking $\frac{1}{2}$ of the numbers $\pm i$ or $\pm 1$ in the parentheses
$(\;\;)$ (nilpotents) and $[\;\;]$ (projectors). 
We look for the  solutions  of  Eq.~(\ref{weylsimpl}) for a particular choice of the $d$-momentum
$p^a=(p^0,0,0,p^3,0,0)$, 
and find what is presented in Table~\ref{Table I.}.
 \begin{table}
 \begin{center}
 \begin{tabular}{|c|c|c|c|c|c|c|}
 \hline
 $\psi^{pos}_{i}$& positive energy state& 
   $\frac{p^0}{|p^{0}|}$&  $\frac{p^3}{|p^{3}|}$&$(-2iS^{03})$&$\Gamma^{(3+1)}$& $S^{56}$
  \\
 \hline
 $\psi^{pos}_{1}$& 
 $ \stackrel{03}{(+i)}\,\stackrel{12}{(+)}|\stackrel{56}{(+)}\,e^{-i|p^0| x^0 + i|p^3| x^3}$& 
  $+1$&  $+1$&$+1$&$+1$ & $\frac{1}{2}$
 \\
 $\psi^{pos}_{2}$& 
  $ \stackrel{03}{(+i)}\,\stackrel{12}{[-]}|\stackrel{56}{[-]}\,e^{-i|p^0| x^0 + i |p^3| x^3}$& 
   $+1$&  $+1$&$+1$&$-1$&$-\frac{1}{2}$
 \\
 $\psi^{pos}_{3}$& 
  $ \stackrel{03}{[-i]}\,\stackrel{12}{[-]}|\stackrel{56}{(+)}\,e^{-i |p^0| x^0 - i|p^3| x^3}$& 
   $+1$&  $-1$&$-1$&$+1$&$\frac{1}{2}$
 \\
 $\psi^{pos}_{4}$& 
  $ \stackrel{03}{[-i]}\,\stackrel{12}{(+)}|\stackrel{56}{[-]}\,e^{-i |p^0| x^0 - i |p^3| x^3}$& 
   $+1$&  $-1$&$-1$&$-1$&$-\frac{1}{2}$
 \\
 \hline
 $\psi^{neg}_{i}$& negative energy state& 
    $\frac{p^0}{|p^{0}|}$&  $\frac{p^3}{|p^{3}|}$&$(-2iS^{03})$&$\Gamma^{(3+1)}$& $S^{56}$
   \\
   \hline
 $\psi^{neg}_{1}$& 
  $ \stackrel{03}{(+i)}\,\stackrel{12}{(+)}|\stackrel{56}{(+)}\,e^{i|p^0| x^0 - i|p^3| x^3}$& 
   $-1$&  $-1$&$+1$&$+1$&$\frac{1}{2}$
 \\
  $\psi^{neg}_{2}$& 
   $ \stackrel{03}{(+i)}\,\stackrel{12}{[-]}|\stackrel{56}{[-]}\,e^{i|p^0| x^0 - i|p^3| x^3}$& 
    $-1$&  $-1$&$+1$&$-1$&$-\frac{1}{2}$
 \\
  $\psi^{neg}_{3}$& 
   $ \stackrel{03}{[-i]}\,\stackrel{12}{[-]}|\stackrel{56}{(+)}\,e^{i|p^0| x^0 + i|p^3| x^3}$& 
    $-1$&  $+1$&$-1$&$+1$&$\frac{1}{2}$
 \\
  $\psi^{neg}_{4}$& 
   $ \stackrel{03}{[-i]}\,\stackrel{12}{(+)}|\stackrel{56}{[-]}\,e^{i|p^0| x^0 + i|p^3| x^3}$& 
    $-1$&  $+1$&$-1$&$-1$&$-\frac{1}{2}$
 \\
 \hline
 \end{tabular}
 \end{center}
 \caption{\label{Table I.} Four positive energy states and four  negative energy states, the 
 solutions of Eq.~(\ref{weylsimpl}), half have $\frac{p^3}{|p^3|}$ positive and half negative.
 $p^a=(p^0,0,0,p^3,0,0)$, 
 $\Gamma^{(5+1)}=-1$, $S^{56}$ defines charges in $d=(3+1)$. Nilpotents 
 $ \stackrel{ab}{(k)}$ and  projectors $ \stackrel{ab}{[k]}$  operate on the vacuum state 
 $|vac>_{fam}$ not written in the table.    
}
 \end{table}
\subsubsection{Solutions of the Weyl equations in $d=(13+1)$}
\label{13+1}

There are $2^{\frac{d}{2}-1}= 64$ basic spinor states of one family representation in $d=(13+1)$.
(We again do not pay attention on the families, since all behave equivalently with respect to 
the discrete symmetries presented in Eqs.~(\ref{calCH}, \ref{calTPH}).)
We present in this subsection  positive energy states for quarks of a particular charge ($\tau^{33}=1/2$, 
$\tau^{38}=1/(2\sqrt{3})$). 
The solution for, say, the 
right handed $u$-quark with spin up, $u_{1\,R}$, with the colour charge ($\tau^{33}=-1/2$ 
and $\tau^{38}=1/(2\sqrt{3})$), weak chargeless and with a positive momentum $p^3$ is proportional to 
($ \stackrel{03}{(+i)}\,\stackrel{12}{(+)}|\stackrel{56}{(+)}\,\stackrel{78}{(+)}||
  \stackrel{9 \;10}{[-]}\;\;\stackrel{11\;12}{[+]}\;\;\stackrel{13\;14}{(-)} 
  \,e^{-ip^0 x^0 + ip^3 x^3}$). 
All the other colour states  follow from  this one 
by the application of the generators  $\tau^{3i}$  of the colour group $SU(3)$, the  
 definition of which, 
 expressed as the superposition of $S^{ab}$, can be found in the caption of
Table~\ref{Table II.}.   One can as well define the generators of the total angular momentum $J^{ab}=
L^{ab} + S^{ab}$. The definition of the generators of the charge groups, presented in the caption of  
Table~\ref{Table II.}, then changes correspondingly by replacing $S^{ab}$ by $J^{ab}$. 
%
 \begin{table}
 \begin{center}
 \begin{tabular}{|c|c|c|c|c|c|c|c|c|c|c|}
 \hline
  $\psi^{pos}_{i}$& positive energy state & 
    $\frac{p^0}{|p^{0}|}$&  $\frac{p^3}{|p^{3}|}$&$(-2iS^{03})$&$\Gamma^{(3+1)}$
  & $\tau^{13}$& $\tau^{23}$& $\tau^{4}$&$ Y$ & $Q$\\
 \hline
 $u_{1\,R}$&
  $ \stackrel{03}{(+i)}\,\stackrel{12}{(+)}|\stackrel{56}{(+)}\,\stackrel{78}{(+)}||
  \stackrel{9 \;10}{(+)}\;\;\stackrel{11\;12}{(-)}\;\;\stackrel{13\;14}{(-)}\, e^{-i|p^0| x^0+i|p^3| x^3}$ & 
  $+1$& $+1$&$+1$& $+1$&$0$& $\frac{1}{2}$&$\frac{1}{6}$& $\frac{2}{3}$&$\frac{2}{3}$\\
$u_{2\,R}$&
  $ \stackrel{03}{[-i]}\,\stackrel{12}{[-]}|\stackrel{56}{(+)}\,\stackrel{78}{(+)}||
  \stackrel{9 \;10}{(+)}\;\;\stackrel{11\;12}{(-)}\;\;\stackrel{13\;14}{(-)}\, e^{-i|p^0| x^0 - i|p^3| x^3}$ & 
  $+1$& $-1$&$-1$& $+1$&$0$& $\frac{1}{2}$&$\frac{1}{6}$& $\frac{2}{3}$&$\frac{2}{3}$\\
 $d_{1\,R}$&
  $ \stackrel{03}{(+i)}\,\stackrel{12}{(+)}|\stackrel{56}{[-]}\,\stackrel{78}{[-]}||
  \stackrel{9 \;10}{(+)}\;\;\stackrel{11\;12}{(-)}\;\;\stackrel{13\;14}{(-)}\, e^{-i|p^0| x^0 + i|p^3| x^3}$ & 
  $+1$& $+1$&$+1$& $+1$&$0$& $-\frac{1}{2}$&$\frac{1}{6}$& $-\frac{1}{3}$&$-\frac{1}{3}$\\
$d_{2\,R}$&
  $ \stackrel{03}{[-i]}\,\stackrel{12}{[-]}|\stackrel{56}{[-]}\,\stackrel{78}{[-]}||
  \stackrel{9 \;10}{(+)}\;\;\stackrel{11\;12}{(-)}\;\;\stackrel{13\;14}{(-)}\, e^{-i|p^0| x^0 - i|p^3| x^3}$ & 
  $+1$& $-1$&$-1$& $+1$&$0$& $-\frac{1}{2}$&$\frac{1}{6}$& $-\frac{1}{3}$&$-\frac{1}{3}$\\
 $d_{1\,L}$&
   $ \stackrel{03}{[-i]}\,\stackrel{12}{(+)}|\stackrel{56}{[-]}\,\stackrel{78}{(+)}||
   \stackrel{9 \;10}{(+)}\;\;\stackrel{11\;12}{(-)}\;\;\stackrel{13\;14}{(-)}\, e^{-i|p^0| x^0-i|p^3| x^3}$ & 
   $+1$& $-1$&$-1$& $-1$& $-\frac{1}{2}$&$0$&$\frac{1}{6}$& $\frac{1}{6}$&$-\frac{1}{3}$\\
 $d_{2\,L}$&
   $ \stackrel{03}{(+i)}\,\stackrel{12}{[-]}|\stackrel{56}{[-]}\,\stackrel{78}{(+)}||
   \stackrel{9 \;10}{(+)}\;\;\stackrel{11\;12}{(-)}\;\;\stackrel{13\;14}{(-)}\, e^{-i|p^0| x^0 + i|p^3| x^3}$ & 
   $+1$& $+1$&$+1$& $-1$& $-\frac{1}{2}$&$0$&$\frac{1}{6}$& $\frac{1}{6}$&$-\frac{1}{3}$\\
  $u_{1\,L}$&
   $ \stackrel{03}{[-i]}\,\stackrel{12}{(+)}|\stackrel{56}{(+)}\,\stackrel{78}{[-]}||
   \stackrel{9 \;10}{(+)}\;\;\stackrel{11\;12}{(-)}\;\;\stackrel{13\;14}{(-)}\, e^{-i|p^0| x^0 - i|p^3| x^3}$ & 
   $+1$& $-1$&$-1$& $-1$& $\frac{1}{2}$&$0$&$\frac{1}{6}$& $\frac{1}{6}$&$\frac{2}{3}$\\
 $u_{2\,L}$&
   $ \stackrel{03}{(+i)}\,\stackrel{12}{[-]}|\stackrel{56}{(+)}\,\stackrel{78}{[-]}||
   \stackrel{9 \;10}{(+)}\;\;\stackrel{11\;12}{(-)}\;\;\stackrel{13\;14}{(-)}\, e^{-i|p^0| x^0 + i|p^3| x^3}$ & 
  $+1$& $+1$&$+1$& $-1$& $\frac{1}{2}$&$0$&$\frac{1}{6}$& $\frac{1}{6}$&$\frac{2}{3}$\\
 \hline
 \end{tabular}
 \end{center}
 \caption{\label{Table II.} One  $SO(7,1)$ sub representation of the representation of $SO(13,1)$, 
 the one representing quarks, which  carry the colour charge ($\tau^{33}=1/2$, $\tau^{38}=1/(2\sqrt{3})$).
 All members have $\Gamma^{(13+1)}=-1$. All  states are the eigenstates of the 
 Cartan subalgebra ($S^{03}, S^{12}, S^{56}, S^{7,8}, S^{9\,10}, S^{11\,12}, S^{13\,14}$) 
 with the eigenvalues defined in Eq.~(\ref{grapheigen}) and solve the 
 Weyl equation~(\ref{weylsimpl}) for the choice of the coordinate system $p^a=(p^0,0,0,p^3,0,\dots,0)$. 
 The infinitesimal generators   of the weak charge $SU(2)$ 
 group are defined as 
 ($\vec{\tau}^1 = \frac{1}{2}(S^{58}- S^{67},  S^{57}+ S^{6,8}, S^{56}- S^{7,8})$), of another $SU(2)$ 
 as  ($\vec{\tau}^2 = \frac{1}{2}(S^{58}+ S^{6,7},  S^{57}- S^{6,8}, S^{56}+ S^{7,8})$), of the $\tau^{4}$ 
 charge as 
 ($- \frac{1}{3}(S^{9\, 10}+  S^{11\, 12} + S^{13\, 14}) $) and of the colour charge group as 
 ($\vec{\tau}^3 = (\frac{1}{2}(S^{9\, 12}- S^{10\,11}, S^{9\, 11}+ S^{10\,12}, S^{ 9\, 10}- S^{11\,12}, 
                S^{9\, 14}- S^{10\,13}, S^{9\, 13}+ S^{10\,14}, S^{11\, 14}- S^{12\,13}, 
               S^{11\, 13}+ S^{12\,14}, \frac{1}{\sqrt{3}}\,(S^{9\, 10}+ S^{11\,12} -2 S^{13\, 14})$),
 $Y = \tau^{23} + \tau^4$, $Q= \tau^{13}+Y$. 
 Nilpotents $ \stackrel{ab}{(k)}$ and  projectors $ \stackrel{ab}{[k]}$  operate on the vacuum state 
 $|vac>_{fam}$ not written in the table.          
}
 \end{table}

\section{Discrete symmetries in $d$ even 
with the desired properties in $d=(3+1)$}
\label{dn}

In section~\ref{dh} we define the discrete symmetries in spaces with $d>(3+1)$ as they follow from 
the definition in $d=(3+1)$. This definition, however, does not allow to interpret the  
angular momentum (the spin, indeed, at the low energy regime) in higher than four 
dimensions as charges in $(3+1)$. The proposed charge conjugated states have, namely, the same 
charges as the starting states. 

We look for new discrete symmetries, which would lead to the desired properties of the anti-particle 
state to any second quantized state:\\
{\bf i. } The anti-particle state has the same momentum in $d=(3+1)$ as the starting state.\\
{\bf ii.} The anti-particle state has the opposite values of the Cartan subalgebra of the 
total angular momentum $J^{st}=L^{st} + S^{st}\,, (s,t)\in (5,6,\dots ,d)$  
(or  at low energies rather the opposite values of the Cartan subalgebra of $S^{st}\,, 
(s,t)\in (5,6,\dots ,d)$) as the starting state.\\

The manifestation of the total angular momentum (in the low energy regime rather the 
spin degrees of freedom) in $d>4$ as charges in $d\le 4$ depends on the 
symmetries that (non-)compact spaces manifest~\cite{DHN}. (For the toy model~\cite{DHN} in 
$d=(5+1)$ the spin on the 
infinite surface, curled into an almost sphere, manifests for a massless spinor as a 
charge in $d=(3+1)$. Only to the massive states the total angular momentum in $d=(5,6)$ contributes.) 
In the case of the {\it spin-charge-family} theory in $d=(13+1)$,  which manifests at low energies  
properties of the {\it standard model}, the 
operators $\vec{\tau}^{1}, 
\vec{\tau}^{2}$, $\vec{\tau}^{3}$, $Y, \tau^{4}, Q$, or rather their superposition (which all are 
superposition of $S^{ab}, 
a,b\in\{5,6,\dots,14\}$) define the conserved charges in $d=(3+1)$ before and after the electroweak 
break.

We define new discrete symmetries by  transforming the above  defined discrete symmetries 
(${\underline {\bf \mathbb{C}}}_{{{\bf \cal H}}}$, $\mathbb{C}_{{ \cal H}}$, ${\cal C}_{{\cal H}}\,$, 
${\cal T}_{{\cal H}}\,,$ ${\cal P}_{{\cal H}}\,$) so that, while 
remaining within the same groups of symmetries, the redefined discrete symmetries manifest the 
experimentally acceptable properties in $d=(3+1)$, which is of the essential importance  for all 
the Kaluza-Klein  theories~\cite{kk,zelenaknjiga,witten,sugra} without  
any degrees of freedom of fermions besides the spin and family quantum numbers~\cite{norma,NPLB}. 
We define new discrete symmetries as follows
\begin{eqnarray}
\label{CNsq}
{\cal C}_{{\cal N}}  &= & {\cal C}_{{\cal H}}
 \, {\cal P}^{(d-1)}_{{\cal H}} \,e^{i \pi J_{1\,2}}\,
e^{i \pi J_{3\,5}}\, e^{i \pi J_{7\,9}}\,e^{i \pi J_{11\,13}},\dots, e^{i \pi J_{(d-3)(d-1)}}\,,\nonumber\\
{\cal T}_{{\cal N}}  &= & {\cal T}_{{\cal H}} \, {\cal P}^{(d-1)}_{{\cal H}} \,e^{i \pi J_{1\,2}}\,
e^{i \pi J_{3\,6}}\, e^{i \pi J_{8\,10}}\,e^{i \pi J_{12\,14}},\dots, e^{i \pi J_{(d-2)d}}\,,\nonumber\\
{\cal P}^{(d-1)}_{{\cal N}}  &= & {\cal P}^{(d-1)}_{{\cal H}} \,e^{i \pi J_{5\,6}}\,
e^{i \pi J_{7\,8}}\, e^{i \pi J_{9\,10}}\,e^{i \pi J_{11\,12}}\,e^{i\pi J_{13\,14}},\dots, e^{i \pi J_{(d-1)d}}\,,
\nonumber\\
\mathbb{C}_{{ \cal N}} &=& 
\mathbb{C}_{{ \cal H}}\, {\cal P}^{(d-1)}_{{\cal H}} \,e^{i \pi J_{1\,2}}\,
e^{i \pi J_{3\,5}}\, e^{i \pi J_{7\,9}}\,e^{i \pi J_{11\,13}},\dots, e^{i \pi J_{(d-3)(d-1)}}\,,
\nonumber\\
{\underline {\bf \mathbb{C}}}_{{{\bf \cal N}}}&=& 
{\underline {\bf \mathbb{C}}}_{{{\bf \cal H}}}
\, {\cal P}^{(d-1)}_{{\cal H}} \,e^{i \pi J_{1\,2}}\,
e^{i \pi J_{3\,5}}\, e^{i \pi J_{7\,9}}\,e^{i \pi J_{11\,13}},\dots, e^{i \pi J_{(d-3)(d-1)}}\,.
\end{eqnarray}
%
The operator for "emptying" is  defined in Eq.(\ref{empt}) as   
$"emptying"= \prod_{\Re \gamma^a}\, \gamma^a \,K $,  
the operator $\mathbb{C}_{{ \cal N}}$ =  $\prod_{\Re \gamma^a}\, \gamma^a \,K $
${\cal C}_{{\cal N}}$, while the operator ${\underline {\bf \mathbb{C}}}_{{{\bf \cal N}}}$ is defined 
according to Eq.~(\ref{makingantip}) as
\begin{eqnarray}
\label{makingantipN}
{\underline {\bf {\Huge \Psi}}}^{\dagger}_{aN}[\Psi^{pos}_{p}]\, |vac>  &=& 
{\underline {\bf \mathbb{C}}}_{{{\bf \cal N}}}\, 
{\underline {\bf {\Huge \Psi}}}^{\dagger}_{p}[\Psi^{pos}_{p}]\, |vac>  =  
\int \, {\mathbf{\Psi}}^{\dagger}_{aN}(\vec{x})\, 
({\bf \mathbb{C}}_{\cal N}\,\Psi^{pos}_{p} (\vec{x})) \,d^{d-1} x  \, \,|vac> \,.
\end{eqnarray}
The rotations ($e^{i \pi J_{1\,2}}\,e^{i \pi J_{3\,5}}\, e^{i \pi J_{7\,9}}\,\dots, e^{i \pi J_{(d-3)(d-1)}})$ 
together  with  (multiplied by) ${\cal P}^{(d-1)}_{{\cal H}}$, which are included in 
${\underline {\bf \mathbb{C}}}_{{{\bf \cal N}}} $ (and in $\mathbb{C}_{{\cal N}}$ and ${\cal C}_{{\cal N}} )$,
keep $p^{i}$  for $i=(1,2,3)$ unchanged,  
while they transform a state so that  all the eigenvalues of the Cartan subalgebra  except 
$S^{03}$ and $J^{12}$ (or at the low energy regime $S^{12}$) change sign~\footnote{Since in our extra 
dimension picture $J_{35}$ is no longer a symmetry (for the metric taken as a background field) in coordinate 
space, the operation $e^{i \pi J_{3\,5}}$ looks suspicious as being
not a symmetry, but it is. Indeed, the operation $e^{i \pi J_{3\,5}}$ is in the coordinate part  composed  
just of a mirror reflection around the $x^3=0$ plane in usual space and reflection in the extra dimension 
space around the surface $x^6=0$. }. 
Correspondingly this redefined ${\underline {\bf \mathbb{C}}}_{{{\bf \cal H}}}$ 
transforms a second quantized state into  the anti-particle state with the same four momentum as the 
starting state but with the opposite values of the total angular momentum (or at the low energy regime 
rather the spin) determined by the Cartan subalgebra eigenvalues, except for $S^{03}$ and $J^{12}$ 
(or at the low energy regime $S^{12}$).
The parity operator $ {\cal P}^{(d-1)}_{{\cal N}} $ changes $p^i$ into $-p^i$ only for $i=(1,2,3)$, 
while the time reversal operator corrects all the properties of the new 
${\underline {\bf \mathbb{C}}}_{{{\bf \cal N}}} $ (and $\mathbb{C}_{{\cal N}}$) and 
${\cal P}^{(d-1)}_{{\cal N}}$ so that
\begin{eqnarray}
\label{CPTNsq}
{\cal C}_{{\cal N}} {\cal P}^{(d-1)}_{{\cal N}} {\cal T}_{{\cal N}} &=&
{\cal C}_{{\cal H}} {\cal P}^{(d-1)}_{{\cal H}} {\cal T}_{{\cal H}} \rightarrow 
\Gamma^{(d)}\, I_{x}\,,\nonumber\\
{\underline {\bf \mathbb{C}}}_{{{\bf \cal N}}} {\cal P}^{(d-1)}_{{\cal N}} {\cal T}_{{\cal N}} &=&
{\underline {\bf \mathbb{C}}}_{{{\bf \cal H}}} {\cal P}^{(d-1)}_{{\cal H}} {\cal T}_{{\cal H}}\,.
\end{eqnarray}
%
%
All three new operators commute among themselves as also the old ones do. 
The shorter expressions for 
the same discrete operators of Eq.~(\ref{CNsq}) are up to a phase 
\begin{eqnarray}
\label{CNsqshorter0}
{\cal C}_{{\cal N}}  &= & \prod_{\Im \gamma^m, m=0}^{3} \gamma^m\,\, \,\Gamma^{(3+1)} \,
K \,I_{x^6,x^8,\dots,x^{d}}  \,,\nonumber\\
{\cal T}_{{\cal N}}  &= & \prod_{\Re \gamma^m, m=1}^{3} \gamma^m \,\,\,\Gamma^{(3+1)}\,K \,
I_{x^0}\,I_{x^5,x^7,\dots,x^{d-1}}\,,\nonumber\\
{\cal P}^{(d-1)}_{{\cal N}}  &= & \gamma^0\,\Gamma^{(3+1)}\, \Gamma^{(d)}\, I_{\vec{x}_{3}}
\,,\nonumber\\
\mathbb{C}_{{ \cal N}} &=& 
\prod_{\Re \gamma^a, a=0}^{d} \gamma^a \,\,\,K
\, \prod_{\Im \gamma^m, m=0}^{3} \gamma^m \,\,\,\Gamma^{(3+1)} \,K \,I_{x^6,x^8,\dots,x^{d}}\nonumber\\
&=& 
\prod_{\Re \gamma^s, s=5}^{d} \gamma^s \, \,I_{x^6,x^8,\dots,x^{d}}\,.
\end{eqnarray}
Operators $I$ operates as follows: $I_{x^5,x^7,\dots,x^{d-1}}$ $(x^0,x^1,x^2,x^3,x^5,x^6,x^7,x^8,
\dots,x^{d-1},x^d)$ $=(x^0,x^1,x^2,x^3,-x^5,x^6,-x^7,\dots,-x^{d-1},x^d)$; $I_{x^6,x^8,\dots,x^d}$ 
$(x^0,x^1,x^2,x^3,x^5,x^6,x^7,x^8,\dots,x^{d-1},x^d)$
$=(x^0,x^1,x^2,x^3,x^5,-x^6,x^7,-x^8,\dots,x^{d-1},-x^d)$, $d=2n$. 

The above defined  operators $\mathbb{C}_{{\cal H}}, {\cal P}^{(d-1)}_{{\cal H}} $ and 
${\cal T}_{{\cal H}}$ (Eqs.~(\ref{calCH}, \ref{calTPH}, \ref{emptmathCPTH})), indexed by ${\cal H}$, 
are good symmetries only when also 
boson fields, in the Kaluza-Klein theories the gravitational fields,   
in higher than $(3+1)$ dimensions are correspondingly transformed and not  considered as {\em background} fields.
However, the operators $\mathbb{C}_{{\cal N}}, {\cal P}^{(d-1)}_{{\cal N}} $ and 
${\cal T}_{{\cal N}}$ with index ${\cal N}$ will be good symmetries even if we take it that 
there is a {\em background} field depending only on the extra dimension coordinates - independent of
whether the extra dimension space is compactified or not - so that they are not transformed.

One can namely
easily see that the transformations of the coordinates of the extra dimensions in 
Eqs.~(\ref{CNsq}, \ref{CNsqshorter0})
are {\em cancelled} between the $\pi$-rotations and the actions of e.g. $P_{\cal H}$ on the extra dimensional
coordinates. Thus it can be easily seen that even if we consider a background gravitational field for the 
extra dimensions - but the $(3+1)$ dimensional space is either flat or  their gravitational field 
is considered dynamical 
so as to be also transformed - these operators with index ${\cal N}$, 
$\mathbb{C}_{{\cal N}}, {\cal P}^{(d-1)}_{{\cal N}} $ and 
$ {\cal T}_{{\cal N}}$, are good symmetries with respect to the space-time transformations.
They  are indeed good symmetries 
according to their action on the Weyl field. 
The crucial point really is that the ${\cal N}$-indexed operators 
$\mathbb{C}_{{\cal N}}, {\cal P}^{(d-1)}_{{\cal N}} $ 
and 
$ \mathbb{T}_{{\cal N}}$ with their associated $x$-transformations do {\em not transform the extra $(d-4)$
coordinates} so that background fields depending on these extra dimension coordinates do not matter.

\subsection{Free spinors}
\label{freespinN}

Let us now see on  two cases, for $d=(5+1)$ and for  $d=(13+1)$,  
how do the new proposals for the discrete symmetries, ${\underline {\bf \mathbb{C}}}_{{{\bf \cal N}}}\,,
\, {\cal P}^{(d-1)}_{{\cal N}},\,{\cal T}_{{\cal N}}$, manifest for non interacting spinors.

{\bf Charge conjugation symmetry ${\underline {\bf \mathbb{C}}}_{{{\bf \cal N}}}$}:

Let us start with $\psi^{pos}_{1}$ from Table~\ref{Table I.}. 
In $d=(5+1)$ the charge conjugation 
operator ${\underline {\bf \mathbb{C}}}_{{{\bf \cal N}}}$ equals to 
${\underline {\bf \mathbb{C}}}_{{{\bf \cal H}}}$ \,$ {\cal P}^{(d-1)}_{{\cal H}} \,$ 
$e^{i \pi J_{12}}\,e^{i \pi J_{35}}$. 
To test this symmetry on the second quantized state 
${\underline {\bf {\Huge \Psi}}}^{\dagger}[\Psi^{pos}_{1}] $ one can start with Eq.~(\ref{calC1}) and 
the recognition below this equation that ${\underline {\bf \mathbb{C}}}_{{{\bf \cal H}}}$ transforms a 
second quantized state 
${\underline {\bf {\Huge \Psi}}}^{\dagger}[\Psi^{pos}_{1}]$  into the anti-particle second quantized 
state with the  properties as the starting state: The same $d$-momentum and 
the same eigenvalues of the Cartan subalgebra operators ($S^{03}$, $J^{12}$, $J^{56}$, or rather 
$S^{12}$, $S^{56}$). 
One can easily check that the operation of ${\cal P}^{(d-1)}_{{\cal H}} \,e^{i \pi J_{12}}\,
e^{i \pi J_{35}}$ on this anti-particle state (the hole in the Dirac sea) with the properties 
$S^{03}=\frac{i}{2}$, $S^{12}=\frac{1}{2}$, $S^{56}=\frac{1}{2}$ and the momentum 
$(|p^0|,0,0,|p^3|,0,0)$ (manifesting in $e^{-ip^0 x^0 +ip^3 x^3}$) transforms this anti-particle 
state into the anti-particle state   $\stackrel{03}{(+i)}\,\stackrel{12}{(+)}|\stackrel{56}{[-]}
\,e^{-ip^0 x^0 + ip^3 x^3}$ put on the top of the Dirac sea, with the same spin and the same handedness 
in $d=(3+1)$ and the opposite "charge": $S^{56}=-\frac{1}{2}$ - 
if we recognize the spin in $d=(5,6)$ as the charge in $d=(3+1)$ - as the starting second 
quantized state. But  ${\cal C}_{{\cal N}} 
\psi^{pos}_{1}= \stackrel{03}{(+i)}\,\stackrel{12}{[-]}|\stackrel{56}{(+)}\,e^{i p^0 x^0 - i p^3 x^3} $ 
(solving the Weyl equation~(\ref{weylsimpl})) does not belong to the same Weyl representation 
as the starting state 
$\Psi^{pos}_{1}$ and also  $\stackrel{03}{(+i)}\,\stackrel{12}{(+)}|\stackrel{56}{[-]}
\,e^{-ip^0 x^0 + ip^3 x^3}$ does not. We can conclude that the charge conjugation operator 
${\underline {\bf \mathbb{C}}}_{{{\bf \cal N}}}$, 
\begin{eqnarray}
\label{CNsq5+1empt}
{\underline {\bf \mathbb{C}}}_{{{\bf \cal N}}} {\underline {\bf {\Huge \Psi}}}^{\dagger}_{p}[\Psi^{pos}_{1}]  
({\underline {\bf \mathbb{C}}}_{{{\bf \cal N}}})^{-1}= 
{\underline {\bf {\Huge \Psi}}}^{\dagger}_{aN}[\mathbb{C}_{{\cal N}}\,\Psi^{pos}_{1}]\,, 
\end{eqnarray}
is not a good symmetry.

Let us make the  charge  conjugation operation ${\underline {\bf \mathbb{C}}}_{{{\bf \cal N}}} $
on  the second quantized state ${\underline {\bf {\Huge \Psi}}}^{\dagger}[{u}_{1 R}]$, the corresponding 
single-particle state of which, put on the top of the Dirac sea, is presented in the first line of 
Table~\ref{Table II.}. 
%
We find in Eq.~(\ref{calC13}) that 
${\cal C}_{{\cal H}} u_{1 R}= \stackrel{03}{(+i)} \stackrel{12}{[-]}| \stackrel{56}{[-]} 
\stackrel{78}{[-]} ||\stackrel{9\,10}{[-]}\stackrel{11\,12}{[+]} \stackrel{13\,14}{[+]}\,
e^{ip^0 x^0 - ip^3 x^3}\,.$ 
To apply  ${\cal C}_{{\cal N}}$ on $ u_{1 R}$  we must, according to the definition in the 
first line of Eq.~(\ref{CNsq}), multiply ${\cal C}_{{\cal H}} u_{1 R}$ by
${\cal P}^{(d-1)}_{{\cal H}} \,e^{i \pi J_{1\,2}}\,
e^{i \pi J_{3\,5}}\, e^{i \pi J_{7\,9}}\,e^{i \pi J_{11\,13}}\,$. We end 
up with
\begin{eqnarray}
\label{CNsq13+1}
{\cal C}_{{\cal N}} u_{1 R} &=& 
\stackrel{03}{(+i)} \stackrel{12}{[-]}| \stackrel{56}{(+)} 
\stackrel{78}{(+)}||\stackrel{9\,10}{(+)}\stackrel{11\,12}{(-)}
 \stackrel{13\,14}{(-)}\,e^{ip^0 x^0 -ip^3 x^3}\,.
\end{eqnarray}
The corresponding second quantized state is the hole in this single particle negative energy state in the 
Dirac sea (Fock space), which solves the Weyl equation for the negative energy state.  
It is the state
\begin{eqnarray}
\label{CNsq13+1empt}
\mathbb{C}_{\cal N}\,u_{1R} &=& 
\stackrel{03}{(+i)} \stackrel{12}{(+)}| \stackrel{56}{[-]} 
\stackrel{78}{[-]}||\stackrel{9\,10}{[-]}\stackrel{11\,12}{[+]}
 \stackrel{13\,14}{[+]}\,e^{-ip^0 x^0 +ip^3 x^3}\,.
\end{eqnarray}
This state, put on the top of the Dirac sea, is the anti-particle state. 
But neither the state of Eq.~(\ref{CNsq13+1}) nor the state of Eq.~(\ref{CNsq13+1empt}) does belong to 
the same Weyl representation, similarly as it was in the case with $d=(5+1)$.
Although the corresponding second quantized state, that is the hole of the state of Eq.~(\ref{CNsq13+1})
in the Dirac sea, which is the same as the state of Eq.~(\ref{CNsq13+1empt}) put on the top of the 
Dirac sea, ${\underline {\bf \mathbb{C}}}_{{{\bf \cal N}}} \,{\bf \underline{u}}_{1R}\, 
({\underline {\bf \mathbb{C}}}_{{{\bf \cal N}}} )^{-1}$ ($\rightarrow 
$ $\stackrel{03}{(+i)} \stackrel{12}{(+)}| \stackrel{56}{[-]} 
\stackrel{78}{[-]}||\stackrel{9\,10}{[-]}\stackrel{11\,12}{[+]}
 \stackrel{13\,14}{[+]}\,e^{-ip^0 x^0 + ip^3 x^3}\,|vac>_{fam}$) has the right charges, that is the 
 opposite ones to 
 those of the corresponding particle state, it is not a good    symmetry.  
Again this is not within the same Weyl representation and correspondingly 
${\underline {\bf \mathbb{C}}}_{{{\bf \cal N}}} $ is not a  good symmetry in $d=(13 +1)$. 

In all the spaces with $d=\,2\,(\mod4)$ the charge conjugation operator 
${\underline {\bf \mathbb{C}}}_{{{\bf \cal N}}} $  is not a good symmetry within one Weyl 
representation: With a product of an  odd number of $\gamma^a$ it jumps out of 
the starting Weyl representation.  

{\bf Parity symmetry ${\cal P}^{(d-1)}_{{\cal N}}$}.

${\cal P}^{(d-1)}_{{\cal N}}$ (the third lines in Eqs.~(\ref{CNsq}, \ref{CNsqshorter0})) 
reflects only in the $d=(3+1)$ and multiplies spinors with $\gamma^0$. It does not keep the 
transformed state within the same Weyl representation, either in the case $d=(5+1)$ or in the case 
$d=(13+1)$. In $d=(5+1)$ it transforms the single particle state $\Psi^{pos}_{1}$ into
$\stackrel{03}{[-i]} \stackrel{12}{(+)}| \stackrel{56}{(+)}
e^{-ip^0 x^0 -ip^3 x^3}\,|vac>_{fam}$, which is not within the same Weyl representation. 
In $d=(13+1)$ ${\cal P}^{(d-1)}_{{\cal N}}$  transforms $u_{1 R}$ into 
$\stackrel{03}{[-i]} \stackrel{12}{(+)}| \stackrel{56}{(+)} 
\stackrel{78}{(+)} ||\stackrel{9\,10}{(+)}\stackrel{11\,12}{(-)} \stackrel{13\,14}{(-)}\,
e^{-ip^0 x^0 -ip^3 x^3}\,|vac>_{fam}\,$, manifesting that ${\cal P}^{(d-1)}_{{\cal N}}$ is 
not a good symmetry in spaces with $d=\,2 \,(\mod 4)$. 

{\bf ${\underline {\bf \mathbb{C}}}_{{{\bf \cal N}}} \times $  ${\cal P}^{(d-1)}_{{\cal N}}$ symmetry}.

Let us now check the ${\underline {\bf \mathbb{C}}}_{{{\bf \cal N}}}\, $$ {\cal P}^{(d-1)}_{{\cal N}}$ 
symmetry. 
According to the third and the fourth line of Eq.~(\ref{CNsq}, \ref{CNsqshorter0})) and to 
Eqs.~(\ref{calCH}, \ref{calTPH}) it contains an even number of $\gamma^a $ operators.
Correspondingly the application of ${\cal C}_{{\cal N}}\, {\cal P}^{(d-1)}_{{\cal N}}$ on any state 
transforms the state again into the state within the same Weyl representation.

In $d=(5+1)$ we apply  ${\underline {\bf \mathbb{C}}}_{{{\bf \cal N}}}\, $$ {\cal P}^{(d-1)}_{{\cal N}}$
 on ${\underline {\bf {\Huge \Psi}}}^{\dagger}_{p}[\Psi^{pos}_{1}]$ by applying 
 $\mathbb{C}_{{\cal N}}\, $$ {\cal P}^{(d-1)}_{{\cal N}}$ on $ \Psi^{pos}_{1}$ as follows: 
 ${\underline {\bf \mathbb{C}}}_{{{\bf \cal N}}}\, $$ {\cal P}^{(d-1)}_{{\cal N}}$ 
 ${\underline {\bf {\Huge \Psi}}}^{\dagger}_{p}[
 \Psi^{pos}_{1}]$ $({\underline {\bf \mathbb{C}}}_{{{\bf \cal N}}}\, $$ {\cal P}^{(d-1)}_{{\cal N}})^{-1} =$ 
 ${\underline {\bf {\Huge \Psi}}}^{\dagger}_{aN}$$[\mathbb{C}_{{\cal N}}$
 ${\cal P}^{(d-1)}_{{\cal N}}\Psi^{pos}_{1}]$.
 One recognizes that it is $\mathbb{C}_{{\cal N}}\, $$ {\cal P}^{(d-1)}_{{\cal N}}$ $ \Psi^{pos}_{1}=$
 $\Psi^{pos}_{4}$ (Table~\ref{Table I.}), 
 which must be put on the top of the Dirac sea, representing the hole in the state $\psi^{neg}_{3}$ 
 in the Dirac sea.
The state  is within the same Weyl and solves the Weyl equation. The 
${\underline {\bf \mathbb{C}}}_{{{\bf \cal N}}}\, $$ {\cal P}^{(d-1)}_{{\cal N}}\,$  
manifests as a good symmetry.

Let in $d=(13+1)$ the operator ${\underline {\bf \mathbb{C}}}_{{{\bf \cal N}}}$
$ {\cal P}^{(d-1)}_{{\cal N}}$
apply on   ${\underline {\bf {\Huge \Psi}}}^{\dagger}_{p}[u_{1 R}]$. One applies correspondingly  
$\mathbb{C}_{{\cal N}}$ ${\cal P}^{(d-1)}_{{\cal N}}$ on $u_{1 R}$, which gives the  state
$\stackrel{03}{[-i]} \stackrel{12}{(+)}| \stackrel{56}{[-]} 
\stackrel{78}{[-]}||\stackrel{9\,10}{[-]}\stackrel{11\,12}{[+]}
\stackrel{13\,14}{[+]}\,e^{-ip^0 x^0 - ip^3 x^3}\,$. This state (which solves the Weyl equation 
$\gamma^a p_{a}\Psi=0$) gives, put on the top of the Dirac sea, 
the corresponding anti-particle, belonging to the same Weyl representation, and it is left handed
with respect $d=(3+1)$. 
%
This anti-particle is recognized as a left handed weak chargeless 
anti $u$-quark, of the anti-colour charge, belonging to the same Weyl representation
(see the ref.~\cite{Portoroz03}, Table 4., line 35).


{\em ${\underline {\bf \mathbb{C}}}_{{{\bf \cal N}}}\, $$ {\cal P}^{(d-1)}_{{\cal N}}$
 is a good symmetry in $d=\,2(2n +1)(=$ $2\,(\mod 4))$
spaces}. 

Following Eq.~(\ref{emptmathCPTH}), the creation operator for an anti-particle state, which is  
${\underline {\bf \mathbb{C}}}_{{{\bf \cal N}}} {\cal P}^{(d-1)}_{\cal N}$  transformed
creation operator for the particle state is therefore
\begin{eqnarray}
\label{emptmathCPN}
{\underline {\bf \mathbb{C}}}_{{{\bf \cal N}}} {\cal P}^{(d-1)}_{\cal N}\,
{\underline {\bf {\Huge \Psi}}}^{\dagger}_{p}[\Psi^{pos}_{1}]\,
({\underline {\bf \mathbb{C}}}_{{{\bf \cal N}}}{\cal P}^{(d-1)}_{\cal N})^{-1}
&=&{\underline {\bf {\Huge \Psi}}}^{\dagger}_{aN}
[\mathbb{C}_{\cal N} {\cal P}^{(d-1)}_{\cal N}\,\Psi^{pos}_{1}]\,.
\end{eqnarray}
$I_{\vec{x}_{3}}$ reflects $(x^1,x^2,x^3)$ and $I_{x^6,x^8,\dots x^d}$ reflects even coordinates in $d>3$.

{\bf Time reversal  ${\cal T}_{{\cal N}}$}.

The application of the time reversal operator ${\cal T}_{{\cal N}}$ (the second equation in 
Eqs.~(\ref{CNsq}, \ref{CNsqshorter0}), constructed in spaces with even $d$ 
out of an even number of $\gamma^a$ operators, 
does keep the transformed state within the same Weyl representation.

Let us test on $d=(5 + 1)$ case first, applying ${\cal T}_{{\cal N}}$ on $\Psi^{pos}_{1}$. 
The transformed state is $\Psi^{pos}_{3}$  from Table~\ref{Table I.}: The state has the same handedness 
in $d=(3+1)$ as the starting state, the same $S^{56}$ eigenvalue and opposite  $p^3$ and $S^{12}$. 
Obviously 
${\cal T}_{{\cal N}}$ is a good symmetry.

In the case of $d=(13 + 1)$ operator ${\cal T}_{{\cal N}}$ transforms $u_{1 R}$ with spin up from 
Table~\ref{Table II.} into the state with spin down ($u_{2 R} = 
\stackrel{03}{[-i]} \stackrel{12}{[-]}| \stackrel{56}{(+)} 
\stackrel{78}{(+)} ||\stackrel{9\,10}{(+)}\stackrel{11\,12}{(-)} \stackrel{13\,14}{(-)}\,
e^{-ip^0 x^0 -ip^3 x^3}\,$), keeping all the  quantum numbers 
except eigenvalue of $S^{03}$ and $S^{12}$ the same  and $p^3$ changes  the sign. The state solves 
the Weyl equation.

 ${\cal T}_{{\cal N}}$ {\em is a good symmetry $d=\, 2\,(\mod 4)$}. 
It  keeps states within the same Weyl representation and 
commutes with the operator $\gamma^a p_a$.

{\bf ${\underline {\bf \mathbb{C}}}_{{{\bf \cal N}}}$ $\times$ ${\cal P}^{(d-1)}_{{\cal N}}\times$  
${\cal T}_{{\cal N}}$ symmetry}.

In $d=(5+1)$ the operator ${\underline {\bf \mathbb{C}}}_{{{\bf \cal N}}} {\cal P}^{(d-1)}_{{\cal N}}\,$ 
${\cal T}_{{\cal N}}$ transforms ${\underline {\bf {\Huge \Psi}}}^{\dagger}_{p}[\Psi^{pos}_{1}]$, with 
$\Psi^{pos}_{1}$ from Table~\ref{Table I.} and creating the particle state, 
into the creation operator for the positive energy anti-particle state 
${\underline {\bf {\Huge \Psi}}}^{\dagger}_{aN}[\Psi^{pos}_{2}]$, since $\mathbb{C}_{ {\cal N}} $
${\cal P}^{(d-1)}_{\cal N}$ ${\cal T}_{{\cal N}}$ $\Psi^{pos}_{1}=$
$\Psi^{pos}_{2}$. This state has an 
opposite handedness in $d=(3+1)$ and also the opposite spin and the opposite "charge".

In $d=(13+1)$ the operator $\mathbb{C}_{{\cal N}}$  ${\cal P}^{(d-1)}_{{\cal N}}$ 
${\cal T}_{{\cal N}}$ transforms the right handed weakless $u_{1\, R}$  quark with spin up 
and colour $(\frac{1}{2},\frac{1}{2 \sqrt{3}})$ from Table~\ref{Table I.}, put on the  top of the Dirac sea, 
into the positive energy anti-particle state with the properties of  $\bar{u}_{1\, L}$ from the 
ref.~\cite{Portoroz03}, 
Table 4., line 36) (put on the top of the Dirac sea): weak chargeless, with the spin down and of the 
anti-colour charge $(-\frac{1}{2},-\frac{1}{2 \sqrt{3}})$.  

${\underline {\bf \mathbb{C}}}_{{{\bf \cal N}}}$ 
${\cal P}^{(d-1)}_{{\cal N}}$ ${\cal T}_{{\cal N}}$ {\em is a good 
symmetry}, as it is expected to be.

\subsection{Interacting spinors}
\label{interspin}

Let us assume quite a general Lagrange density for a spinor in $d=((d-1) +1)$ dimensional space, which 
carries, like in the Kaluza-Klein theories, the spins and no charges 
%
\begin{eqnarray}
\label{lagrangespinsfamilies0}
{\cal L} &=&\frac{1}{2} \,E \, \Psi^{\dagger} \, \gamma^0 \, \gamma^a\, p_{0a}\,\Psi + \, h.c.\,,\nonumber\\
p_{0a}   &=& f^{\alpha}_{a} p_{\alpha} + \frac{1}{2E} \{p_{\alpha},f^{\alpha}_{a} E \}_{-}-
\frac{1}{2}\, S^{cd}\, f^{\alpha}_{a}\,\omega_{cd \alpha}\,. 
\end{eqnarray}
$f^{\alpha}_{a}$ are vielbein and $\omega_{cd \alpha}$ spin connection fields,  the gauge fields of  
$p^{a}$ and $S^{ab}$, respectively. In this paper we do not discuss the families quantum numbers, 
which commute with here defined discrete symmetries operators.
%
%
%
%
Let the vielbeins and spin connections be responsible for the break of symmetry of $M^{(d-1)+1}$ into
$M^{3+1}\times$ $M^{d-4}$ so that the manifold $M^{d-4}$ is (almost) compactified and let the spinor 
manifest in $d=(3+1)$ the ordinary spin and the charges
~\footnote{In the references~\cite{NH} it is demonstrated on the toy model how such an almost 
compactification could occur.}.
Looking for the subgroups (denoted by $B,C$)  of the  $SO((d-1)+1)$ 
group and assuming no gravity in $d=(3+1)$, the Lagrange density of Eq.~(\ref{lagrangespinsfamilies0}) 
can be rewritten in a more familiar shape 
\begin{eqnarray}
\label{lagrangespinsfamilies1}
{\cal L} &=& \frac{1}{2} \,E \, \Psi^{\dagger} \, \gamma^0 \, (\gamma^m \, p_{0m} 
+ \gamma^s \, p_{0s})\,\,\Psi \,+ h.c. \,,\nonumber\\
p_{0m}   &=&  p_{m} - \sum_{B}\,\vec{\tau}^{B}\,\vec{A}^{B}_{m}\,,\nonumber\\
p_{0s}   &=&  f^{\sigma}_{s} \,p_{\sigma} + \frac{1}{2E} \{p_{\sigma},f^{\sigma}_{s} E \}_{-}
- \sum_{C}\, \vec{\tau}^{C}\,\vec{A}^{C}_{s}\,,
\end{eqnarray}
with $m=(0,1,2,3)$,  $s=(5,6,\dots,d)$. We have 
$\tau^{Bi}=        \sum_{st}\, b^{Bi}{}_{st}\, S^{st}\,$, 
$\tau^{Ci}=        \sum_{st}\, c^{Ci}{}_{st}\, S^{st}\,$, 
$\sum_{B}\,\vec{\tau}^{B}\,\vec{A}^{B}_{m}= \frac{1}{2}\, \sum_{s,t}\,S^{st}\, \omega_{st m}\,$,
$\sum_{C}\,\vec{\tau}^{C}\,\vec{A}^{C}_{s}= \frac{1}{2}\,\sum{st}\, S^{st}\,f^{\sigma}_{s} 
\omega_{st \sigma}\,$. 
%

One finds  that
\begin{eqnarray}
\label{CNtransempt}
\mathbb{C}_{\cal N}\,\,  \tau^{Ai}\,\, \mathbb{C}_{\cal N}^{-1} &=& - \tau^{Ai}\,, \nonumber\\
\mathbb{C}_{\cal N}\, \, A^{Ai}_{m}(x^0,\vec{x}_3)\,\, \mathbb{C}_{\cal N}^{-1} 
&=& -A^{Ai}_{m}(x^0,\vec{x}_3)\,, \nonumber\\
\mathbb{C}_{\cal N}{\cal P}^{(d-1)}_{\cal N}\,\,  \tau^{Ai}\,\, (\mathbb{C}_{\cal N}
{\cal P}^{(d-1)}_{\cal N})^{-1} &=& - \tau^{Ai}\,, \nonumber\\
\mathbb{C}_{\cal N}{\cal P}^{(d-1)}_{\cal N}\,\,  A^{Ai}_{m}(x^0,\vec{x}_3) \,\,(\mathbb{C}_{\cal N}
{\cal P}^{(d-1)}_{\cal N})^{-1} &=& - A^{Ai\,m}(x^0,-\vec{x}_3)\,,\nonumber\\
\mathbb{C}_{\cal N} {\cal P}^{(d-1)}_{\cal N} {\cal T}_{\cal N}\,\, \tau^{Bi}\,A^{Bi}_{m} (x)\,\,
(\mathbb{C}_{\cal N} {\cal P}^{(d-1)}_{\cal N} {\cal T}_{\cal N})^{-1}  
&=& (-\tau^{Bi})\,(-A^{Bi*}_{m}(-x))\,,
\end{eqnarray}
all in agreement with the standard knowledge for the gauge vector fields and charges in $d=(3+1)$~\cite{ItZu}.

One  can  check also that
%
$\mathbb{C}_{\cal N} {\cal P}^{(d-1)}_{\cal N} {\cal T}_{\cal N}\,$ $ \gamma^a$
$(\mathbb{C}_{\cal N} {\cal P}^{(d-1)}_{\cal N} {\cal T}_{\cal N})^{-1}= 
\gamma^a \,$; 
$\mathbb{C}_{\cal N} {\cal P}^{(d-1)}_{\cal N} {\cal T}_{\cal N} $ $ S^{ab}\,\,$
$(\mathbb{C}_{\cal N} {\cal P}^{(d-1)}_{\cal N} {\cal T}_{\cal N})^{-1}=$ $- S^{ab}\,$;
$\mathbb{C}_{\cal N} {\cal P}^{(d-1)}_{\cal N} {\cal T}_{\cal N}$ $f^{\alpha}_{a}(x)\,p_{\alpha}$
$(\mathbb{C}_{\cal N} {\cal P}_{\cal N} {\cal T}_{\cal N})^{-1}$ $ = f^{\alpha\,*}_{a}(-x)\,p_{\alpha}\,$;
$\mathbb{C}_{\cal N} {\cal P}^{(d-1)}_{\cal N} {\cal T}_{\cal N} $ $\omega_{abc}(x)\,\,$
$(\mathbb{C}_{\cal N} {\cal P}^{(d-1)}_{\cal N} {\cal T}_{\cal N})^{-1} $ 
$= - \omega^{*}_{abc}(-x) \,$. 
%
%
We also have 
$\mathbb{C}_{\cal N} {\cal P}^{(d-1)}_{\cal N} {\cal T}_{\cal N}\,\tau^{Ci}\,A^{Ci}_{s} (x)
(\mathbb{C}_{\cal N} {\cal P}^{(d-1)}_{\cal N} {\cal T}_{\cal N})^{-1} $ 
$= (-\tau^{Ci})\,(-A^{Ci*}_{s})(-x))\,$,
%
concerning in $d=(3+1)$ the gauge scalar fields. The later determine  massless and massive solutions
for spinors and, if gaining nonzero vacuum expectation values, contribute not only  to masses of 
spinors but also to those gauge fields, to which they couple. 

There exist in (almost) compactified spaces ${\cal M}^{d-4}$, for  particular choices of vielbeins and 
spin connection fields in Eq.~(\ref{lagrangespinsfamilies0}), massless and massive solutions~\cite{NH,DHN}. 
In subsect.~\ref{comp5+1} we discuss such a case for $d=(5+1)$. 
One finds that the operator $\mathbb{C}_{{ \cal N}} {\cal P}^{(d-1)}_{{ \cal N}}$ transforms either the
massless or massive solutions of the Weyl equation, which represent particle states on the top of 
the Dirac sea, into their anti-particle states, which are holes in the Dirac sea. It follows also for 
the case that the infinite surface in the fifth and the sixth dimensions compactifies into an 
almost $S^2$ with the radius $\rho_{0}$ that the massive state $\psi^{(6)(m \rho_0)}_{(n+ \frac{1}{2})}$
$ = ({\cal A}_{n}\, \stackrel{56}{(+)}\, \psi^{(4)}_{(+)}(\vec{p})  
+ {\cal B}_{n+1}\,e^{i \phi}  \stackrel{56}{[-]}\, \psi^{(4)}_{(-)}(\vec{p}))\, e^{in \phi}$
($\psi^{(4)}_{(\pm)}(\vec{p})$ are the corresponding plane wave solutions in $d=(3+1)$ with the 
three momentum $\vec{p}$) with the charge $(n+\frac{1}{2})$  
($M^{56}$ $\psi^{(6)(m \rho_0)}_{(n+ \frac{1}{2})}$ $=(n+\frac{1}{2})$
$\psi^{(6)(m \rho_0)}_{(n+ \frac{1}{2})}$, $M^{56}$ is the total angular momentum)   
transforms under $\mathbb{C}_{{ \cal N}} {\cal P}^{(d-1)}_{{ \cal N}}$ into 
$\psi^{(6)(m \rho_0)}_{-(n+ \frac{1}{2})}= $ $
({\cal B}_{n+1}\, \stackrel{56}{(+)}\, \psi^{(4)}_{(+)}(-\vec{p})  
+ {\cal A}_{n}\, \, e^{i\phi}\, \stackrel{56}{[-]}\, \psi^{(4)}_{(-)}(-\vec{p}))\, e^{-i(n+1) \phi}$, which is the 
hole in the Dirac sea. This state $\psi^{(6)(m \rho_0)}_{-(n+ \frac{1}{2})}$
solves the by  $\mathbb{C}_{{ \cal N}} {\cal P}^{(d-1)}_{{ \cal N}}$ transformed Weyl equation 
(\ref{equationm56gen1}) with $F \rightarrow -F$ and (${\cal B}_{n+1}= {\cal A}_{-n-1}$, 
${\cal A}_{n}= {\cal B}_{-n}$), as one can check in Eq.~(\ref{equationm56gen1CP}).   

%

The Hermiticity requirement for the Lagrange density (Eq.(\ref{lagrangespinsfamilies1})),
${\cal L}^{\dagger}= {\cal L}$, leads to
\begin{eqnarray}
\label{hcreq}
\omega^{*}_{abc}(x)&=& (\mp)\,\omega_{abc}(x)\,;(-)\, {\rm if}\, a=c \,{\rm or}\, b=c\,, 
(+)\, {\rm otherwise}\,,
\end{eqnarray}
which is to be taken into account together with the $\mathbb{C}_{\cal N} {\cal P}_{\cal N} {\cal T}_{\cal N}$ 
invariance.

\section{Discussions on generality of our proposal for discrete symmetries}
\label{understanding}

Searching for the appropriate definition of the discrete symmetries, when starting at higher dimensions 
than $4$, for any  even $d$, we indeed limited ourselves to a  simple example with two different choices of 
dimension. We assumed that there is a central point symmetry (there might be several) and particular rotational 
symmetries around the central point. We do not really study how could an almost compactification occur. 
There are several papers in the literature~\cite{CEZ2013,BRG} studying the way of compactification. It is not 
easy to say whether the experiences from this literature can usefully be used in our cases. In one of our 
cases we so far just use the appropriate gauge fields, zweibeins and spin connections, without paying 
attention where do they originate and then study the properties of the scalar and vector gauge fields and 
spinors. We shortly present their properties in the subsections of this section.

%
Let us ask first how general is our proposal for Kaluza-Klein type of theories. Although for examples like the 
one when dimensions are compactified into a (compact~\footnote{An almost from the infinite surface 
compactified torus has no conserved charges.}) torus with momenta as the conserved charges would not be of 
our type, still our proposal might be of a help to find the definition of appropriate symmetries also for 
such cases.  

We got the proposals  for the discrete symmetries  for the effective $(3+1)$ theory (Eqs.~(\ref{CNsq}, 
\ref{CNsqshorter0})) from analysing our special case, for which one  immediately sees that the proposal 
for ${\cal{P}}^{d-1}_{\cal{N}}$ does not contain any transformation of the extra dimension coordinates, 
while getting the contribution of the $\gamma^{a}$ 
matrices adjusted so that the extra dimensional gamma matrices $\gamma^5$, $\gamma^6$, ..., $\gamma^{d-1}$,
$\gamma^d$ commute with ${\cal{P}}^{d-1}_{\cal{N}}$. This means that this operation is quite insensitive to 
the extra dimensions in such a way that it is not important if the extra dimensional space obeys any parity 
like symmetry. Correspondingly  there should be no transformation of the extra dimension boson fields in the 
sense that the extra dimensional components should not be changed except for their transformation due to the 
coordinate flipping in  the first three dimensions.
The components with vector or tensor indices among the first three spatial components bring correspondingly 
the signs shifted, but otherwise the boson fields are not to be transformed under ${\cal{P}}^{d-1}_{\cal{N}}$. 
This means successively that the general type of background fields describing the extra dimensional curling 
up in some way will not be modified under this operation and thus one takes such fields as background fields  
in the sense of  this  ${\cal{N}}$-marked parity operation ${\cal{P}}^{d-1}_{\cal{N}}$, which means that one  
leaves such background fields untouched.

Looking at Eq.~(\ref{CNsqshorter0})  one sees that the background fields have to obey some reflection 
symmetry in order that ${\cal{T}}_{\cal{N}}$ and ${\cal{C}}_{\cal{N}}$ be well defined symmetries. 
(One needs well defined 
discrete symmetries even if in particular cases each of them is not a good symmetry, when the 
handedness of spinors prevent them to be a good symmetries, while the product of the two is then a good 
symmetry.) 

So, unless the extra dimensional back ground fields obey in even $d$ the reflection symmetry for 
${\cal C}_{{\cal N}}$ 
\begin{eqnarray}
\label{simplifiescptnC}
 (x^5,x^6,\cdots, x^d) \rightarrow (x^5, -x^6, x^7, -x^8,\cdots, x^{d-1},-x^d) \, ,
\end{eqnarray}
while for ${\cal T}_{{\cal N}}$ they obey
\begin{eqnarray}
\label{simplifiescptnT}
 (x^5,x^6,\cdots, x^d) \rightarrow (-x^5, x^6, -x^7, x^8,\cdots, -x^{d-1},x^d)\, ,
\end{eqnarray}
the equations of motion for spinors  do not have these symmetries of ${\cal{T}}_{\cal{N}}$ and 
${\cal{C}}_{\cal{N}}$. One easily checks that the toy model~\cite{DHN} has the above 
(Eqs.~(\ref{CNsqshorter0}, \ref{simplifiescptnC}, \ref{simplifiescptnT})) symmetry.

These requirements for the extra dimensional reflection for background and fermion fields of 
Eqs.~(\ref{simplifiescptnC},
\ref{simplifiescptnT}) are due to our request that anti-particles should manifest in $(3+1)$ 
dimensions opposite charges as particles (the charges of which correspond to appropriate "Killing forms"). 
(So that $\mathbb{C}_{{\cal N}}$ inverts the charges.)  One can understand the alternating reflection 
properties of $x^s, \, s\ge 5$, Eq.~(\ref{simplifiescptnC}), in our example of the toy model~\cite{DHN}, 
by the requirement that the "Killing forms", which are circles around the fixed point, must change the 
orientation. 

Concerning the alternating reflection (in coordinate space) of ${\cal T}_{{\cal N}}$ in 
Eq.~(\ref{simplifiescptnT}) one 
can understand this alternation by again looking at our example of the toy model~\cite{DHN}. Since  
${\cal T}_{\cal{H}}$ (Eq.~(\ref{calTPH})) reflects the momentum $\vec{p}$ in $(d-1)$ dimensions,  the "Killing 
forms" acquire a change in the direction. To compensate the change of the sign of the "Killing forms"  we need 
the alternative reflection offered by  ${\cal T}_{{\cal N}}$. 
In this way one namely obtains the usually wanted property of the $(3+1)$-dimensional time reversal operator 
${\cal T}_{{\cal N}}$ that it leaves the charges untouched. 

While ${\cal T}_{{\cal H}}$ does change the signs of "Killing forms", $\mathbb{C}_{{\cal H}}$ does not. 
So, both, $\mathbb{C}_{{\cal H}}$ and ${\cal T}_{{\cal H}}$ are cured by the same reflection of "Killing forms":

In an example, when compactification is made by  a torus (let us say again that almost  compactified 
torus has no rotational symmetry), where  the generators of translations around the 
torus are declared as charges in $(3+1)$, we  must replace the reflection symmetry of 
Eqs.~(\ref{simplifiescptnC}, \ref{simplifiescptnT}) by the reflection which again inverts the corresponding 
"Killing forms". This means that $x^s$ goes to $-x^s$, $s=5,6,\dots$ around any point. 

In the torus case we need the true parity  ${\cal P}_{{\cal H}} \times {\cal P}_{{\cal N}}$ in extra dimensions 
to change the signs of "Killing forms". 

In complicated cases we can a priori imagine that constructing appropriate reflections inverting the signs of 
all the to be  charges "Killing forms" could be complicated. 

If the background fields are mainly just the metric tensor fields with extra dimensional components and the 
charges commute, it would not be difficult to find for each separate charge a reflection symmetry, reflecting 
just that symmetry,  just that charge. Combining these reflections for the separate charges to a combined
reflection reflecting all the charges would then be a proposal for the replacement for 
(\ref{simplifiescptnT}) and (\ref{simplifiescptnC}).

Let us mention the ref.~\cite{Colinbook} with one of the authors of this paper (H.B.N.) as a coauthor. 
The book  stresses that symmetries can often be derived from small assumptions which we put into a theory. 
For the discrete symmetries for the strong and electromagnetic interactions one ought to assume: 
i. Anomaly cancellations, ii. Small group representations and iii. Charge  quantization rule. This 
author understands their derivation as a competitive way of deriving the discrete 
symmetries operators without knowing the theory behind. 

Let us add that  the Calabi-Yau kind of spaces~\cite{BRG} seems to have the symmetry  
so that our proposed discrete symmetries work.

\subsection{Comments on two special cases }
\label{comp5+1}

In the subsection~\ref{interspin} we discuss how do our proposed discrete symmetries, 
Eq.~(\ref{CNsq}, \ref{CNsqshorter0}),  behave in cases when there are the vielbein and spin 
connection  fields (Eq.~(\ref{lagrangespinsfamilies0})) of the Kaluza-Klein kinds, which  
determine the spinor interactions. We demonstrate there how do spinors manifest in 
$d=(3+1)$ the Kaluza-Klein charges, interact with the Kaluza-Klein vector  gauge fields 
and with the scalar gauge fields (these last ones determine masses of spinors in $(3+1)$  
and, after gaining nonzero vacuum expectation values, besides the masses of spinors also  
the masses of those  vector gauge fields which they interact with) and how do spinors, 
vector gauge fields and scalar gauge fields transform under our proposed discrete symmetries. 

In  this subsection we shortly present the fields, zweibeins and spin connections, which in 
our toy model~\cite{NH,DHN} in $d=(5+1)$ cause an almost compactification. We also comment 
briefly  our "realistic case" in $d=(13+1)$ which is offering the explanation for all the 
charges and gauge fields of the {\it standard model}, with the families and scalar fields 
included, although we do not discuss in this paper the appearance of families and 
correspondingly a possible explanation for the Yukawa couplings~\cite{NJMP}.

{\bf A toy model in $d=(5+1)$}.

In the ref.~\cite{DHN} we present the zweibeins and the spin connection fields, assumed to 
be caused by a kind of spinor condensates, which allow after the  compactification of the 
manifold $M^{5+1}$ into $M^{3+1} \times$ an almost $S^{2}$
one massless and mass protected solution  and the chain of massive solutions of the Weyl equation
following from the  Lagrange density in Eq.~(\ref{lagrangespinsfamilies0}).  
%
%
We assume a flat $(3+1)$ space and 
the zweibein in $d=(5,6)$
\begin{eqnarray}
e^{s}{}_{\sigma} = f^{-1}
\begin{pmatrix}1  & 0 \\
 0 & 1 
 \end{pmatrix},
f^{\sigma}{}_{s} = f
\begin{pmatrix}1 & 0 \\
0 & 1 \\
\end{pmatrix},
\label{fzwei}
\end{eqnarray}
with 
\begin{eqnarray}
\label{f}
f &=& 1+ (\frac{\rho}{2 \rho_0})^2= \frac{2}{1+\cos \vartheta}\,,\nonumber\\ 
x^{(5)} &=& \rho \,\cos \phi,\quad  x^{(6)} = \rho \,\sin \phi\,, \nonumber\\
E &=& \det(e^s{\!}_{\sigma})=f^{-2}\,, e^s{\!}_{\sigma}\,f^{\sigma}{\!}_{t}= \delta^{s}_{t}\,,
\end{eqnarray}
and the spin connection field 
\begin{eqnarray}
  f^{\sigma}{}_{s'}\, \omega_{st \sigma} &=& i F\, f \, \varepsilon_{st}\; 
  \frac{e_{s' \sigma} x^{\sigma}}{(\rho_0)^2}\, , \quad 
 0 <2F \le 1\, 
  ,\quad s=5,6,\,\,\; \sigma=(5),(6)\,, 
\label{omegas}
\end{eqnarray}
where $\rho_0$ is the radius of $S^2$.
It follows that this choice of the spin connection field on an almost $S^{2}$ allows for $0< 2F \le 1$   
only one normalizable (square integrable) massless solution - 
the left handed spinor with the Kaluza-Klein charge  in $d=(3+1)$ equal to $\frac{1}{2}$.  
The massless and massive  solutions preserve the rotational symmetry around the axis perpendicular to the 
surface in the fifth and the sixth dimension and are correspondingly the eigenfunction of the total 
angular momentum $M^{56}= x^5 p^6-x^6 p^5  + S^{56}= -i \frac{\partial}{\partial \phi} + S^{56}$,
$M^{56}\psi^{(6)}= (n+\frac{1}{2})\,\psi^{(6)}$.
For the choice of the coordinate system $p^a= (p^0,0,0,p^3,p^5,p^6)$ the massive solution with the
Kaluza-Klein charge ${n+1/2}$
\begin{eqnarray}
\psi^{(6)(\rho_0 m) }_{n+1/2}= ({\cal A}_{n}\,\stackrel{03}{(+i)} \stackrel{12}{(+)} \stackrel{56}{(+)}  
+ {\cal B}_{n+1}\, e^{i \phi}\, \stackrel{03}{[-i]}\stackrel{12}{(+)}\,\stackrel{56}{[-]})\,\cdot e^{in \phi}
e^{-i(p^0 x^0- p^3 x^3)}\,,
\label{mabpsi}
\end{eqnarray}
solves  the  equation of motion, derived from the Lagrange function~Eq.(\ref{lagrangespinsfamilies0}), 
with ${\cal A}_{n}$ and ${\cal B}_{n+1}$ determined by the equations
\begin{eqnarray}
&&-if \,\{ \,(\frac{\partial}{\partial \rho} + \frac{n+1}{\rho})  -   
  \frac{1}{2\, f} \, \frac{\partial f}{\partial \rho}\, (1+ 2F)\}  {\cal B}_{n+1} + m {\cal A}_n = 0\,,  
\nonumber\\
&&-if \,\{ \,(\frac{\partial}{\partial \rho} - \quad \frac{n}{\rho}) -   
  \frac{1}{2\, f} \, \frac{\partial f}{\partial \rho}\, (1- 2F)\}  {\cal A}_{n} + m {\cal B}_{n+1} = 0\,.
\label{equationm56gen1}
\end{eqnarray}
There exists the massless left handed spinor with the Kaluza-Klein charge  in $d=(3+1)$ equal to 
$\frac{1}{2}$
\begin{eqnarray}
\psi^{(6)(m=0)}_{\frac{1}{2}} ={\cal N}_0  \; f^{-F+1/2} 
\stackrel{03}{(+i)}\stackrel{12}{(+)}\stackrel{56}{(+)}\,e^{-i(p^0 x^0-p^3x^3)}\,. 
\label{Massless}
\end{eqnarray} 
For $F=\frac{1}{2}$ and  $p^1=0=p^2$ this solution corresponds to the particle 
described by $\psi^{pos}_{1}$ and put on the top of the Dirac sea. The corresponding 
$\mathbb{C}_{\cal N} {\cal P}^{(d-1)}_{\cal N}$ transformed state, put on the top of the Dirac sea, 
that is the anti-particle state,  the hole indeed in the Dirac sea, 
is the state $\psi^{pos}_{4}$ corresponding to the empty $\psi^{neg}_{3}$ 
in the Dirac sea,   in accordance with what we have discussed in sect.~\ref{dn}. With the operator 
$\mathbb{C}_{\cal N} {\cal P}^{(d-1)}_{\cal N}$ transformed state $\psi^{(6)(\rho_0 m) }_{n+1/2}$
is the state 
\begin{eqnarray}
\psi^{(6)(\rho_0 m) }_{-(n+1/2)}= ({\cal A}_{-(n+1)}\,\stackrel{03}{(+i)} \stackrel{12}{(+)} \stackrel{56}{(+)}  
+ {\cal B}_{-n}\, e^{i \phi}\, \stackrel{03}{[-i]}\stackrel{12}{(+)}\,\stackrel{56}{[-]})\,\cdot e^{-i(n+1) \phi}
e^{-i(p^0 x^0+ p^3 x^3)}\,,
\label{mabpsianti}
\end{eqnarray}
with the two functions ${\cal A}_{-(n+1)}$ and ${\cal B}_{-n}$, which solve the equations
\begin{eqnarray}
&&-if \,\{ \,(\frac{\partial}{\partial \rho} - \frac{n}{\rho})  -   
  \frac{1}{2\, f} \, \frac{\partial f}{\partial \rho}\, (1- 2F)\}  {\cal B}_{-n} + m {\cal A}_{-(n+1)} = 0,  
\nonumber\\
&&-if \,\{ \,(\frac{\partial}{\partial \rho} + \quad \frac{n+1}{\rho}) -   
  \frac{1}{2\, f} \, \frac{\partial f}{\partial \rho}\, (1+ 2F)\}  {\cal A}_{-(n+1)} + m {\cal B}_{-n} = 0\,,
\label{equationm56gen1CP}
\end{eqnarray}
where $F$ goes to $-F$, in accordance with the $\mathbb{C}_{{ \cal N}} {\cal P}^{(d-1)}_{{ \cal N}}=$ 
$\gamma^0\, \gamma^5\, I_{\vec{x}_{3}}\, I_{x_6}$ transformation requirement for the fields. 

One easily sees that $\psi^{(6)(\rho_0 m) }_{-(n+1/2)}=-$ $\mathbb{C}_{\cal N} {\cal P}^{(d-1)}_{\cal N}$
$\psi^{(6)(\rho_0 m) }_{(n+1/2)}$.

Would the scalar (with respect to ($d=(3+1)$))  $f^{\sigma}_{s}\omega_{56 \sigma}$ achieve 
nonzero vacuum expectation values breaking the rotational symmetry on the $(5,6)$ surface, 
the charge $S^{56}$ would no longer be conserved and the scalar fields would behave similar as the
Higgs of the {\it standard model}, carrying in this case the "hypercharge" $S^{56}$.

{\bf The case with $d=(13+1)$}.

In the case of $d=(13+1)$ the  compactification is again assumed to be triggered by spinor condensates 
which then cause the appearance of vielbeins and spin connection fields. The compactification 
from the symmetry $SO(13,1)$ (first to $SO(7,1)\times U(1)_{II} \times SU(3)$ and then) to 
$SO(3,1)\times SU(2)_I \times SU(2)_{II}$ $\times U(1)_{II}$ $\times SU(3)$, 
leaving all the family members massless  
(in the toy model case we found the solution for the compactification of the $(x_5,x_6)$ surface into 
an almost $S^2$ for  particular spin connections and vielbeins) ensure that the spins in $d>4$ (in 
the low energy limit, otherwise the total angular momenta) manifest in $d=(3+1)$ all the observed 
charges. (There are in the theory~\cite{NJMP,norma,Portoroz03,pikanorma,gmdn,gn,BledAN2010,NPLB}  
two kinds of  spin connection fields. The second one, not discussed in this paper, takes care of families. 
Correspondingly there are before the electroweak break four, rather than three so far observed,  
massless families of quarks and leptons.)

We don't yet have the solution for the compactification procedure not even comparable with the one for 
the toy model in $d=(5+1)$. This study is under consideration. 

However, analysing a massless left handed representation in $d=(13+1)$ - similarly as in the case of 
the toy model but in this case taking into account the charge groups of quarks and leptons assumed by 
the {\it standard model}, they are subgroups of $SO(13,1)$ - one easily sees that one (each) family 
representation in $d=(13+1)$ contains~\cite{norma,Portoroz03,pikanorma} the left handed (with respect 
to $d=(3+1)$)  weak charged coloured quarks and colourless leptons with particular spinor quantum 
number ($\frac{1}{6}$ for quarks and $-\frac{1}{2}$ for leptons) and zero hyper charge and the right 
handed weak chargeless quarks and leptons, with the spinor charge of the left handed ones but with 
the hyper charges as required by the {\it standard model}. In 
Table~\ref{Table II.} are $u$ and $d$ quarks of a particular colour presented, left and right handed 
ones. Leptons distinguish from the quarks in the colour and in the spinor quantum numbers. One can find 
the whole one family representation in the ref.~\cite{Portoroz03}  and in Table~\ref{Table so13+1.} 
of Appendix~\ref{technique}.

When the scalar spin connection fields of the two kinds (bringing appropriate weak and hyper charges to 
the right handed members of one family) gain nonzero vacuum expectation values, the electroweak break 
occurs, causing that the fermions and the weak bosons become massive, while the $U(1)$ electromagnetic 
field  stay massless. 

The effective Lagrange density  is presented in Eq.(~\ref{lagrangespinsfamilies1}).

 The term  $\bar{\psi} \gamma^{s} p_{0s} \; \psi$ is responsible for  
 masses of spinors in $d=(3+1)$, with $\gamma^0 \gamma^s \,, s=(7,8)$  transforming the right handed 
 quarks and leptons, weak chargeless and of particular hypercharge 
 into the left handed weak charged partners.

Similarly as in the case of the toy model the discrete symmetries of Eq.(~\ref{CNsqshorter0}) keep their 
meaning also in this  case.

\section{Conclusions}
\label{discCPT}

We define in this paper the discrete symmetries, ${\underline {\bf \mathbb{C}}}_{{{\bf \cal N}}}$, 
${\cal P}_{{\cal N}}$  
and ${\cal T}_{{\cal N}}$ (Eqs.(\ref{CNsq}, \ref{CNsqshorter0})) in even dimensional spaces 
leading in $d=(3+1)$ to the experimentally observed symmetries, if the Kaluza-Klein kind of a 
theory~\cite{kk,zelenaknjiga,witten,sugra} with $d> (3+1)$ determining charges in $d=(3+1)$ 
(among them also the {\it spin-charge-family} proposal  
of one of us (N.S.M.B.\cite{NJMP,norma,Portoroz03,pikanorma,gmdn,gn,BledAN2010,norma93}  
offering also the mechanism for generating families), is the right way to 
understand the assumptions of the {\it standard model}. We indeed define three kinds of the 
charge conjugation operators: Besides ${\underline {\bf \mathbb{C}}}_{{{\bf \cal N}}}$, which 
operates on the creation operator for a particle, also ${\cal C}_{\cal N}$ transforming the 
positive energy state representing a particle when put on the top of the Dirac sea into its 
negative energy partner, and $\mathbb{C}_{\cal N}$ which empties this negative energy state in the 
Dirac sea representing on the top of the Dirac sea the anti-particle state~(\ref{makingantipN}).

Although we designed this discrete symmetry 
operators for cases with a central point symmetry (see sect. \ref{understanding}) (there might be several) 
and particular rotational symmetries around the central point in higher dimensions, yet our proposal 
might help to define these discrete symmetries also in more complicated cases, as discussed in 
sect.~\ref{understanding}.

Our  (${\underline {\bf \mathbb{C}}}_{{{\bf \cal N}}}$,
${\cal P}_{{\cal N}}$, ${\cal T}_{{\cal N}}$) discrete symmetries are rotated and reflected with 
respect to the symmetries  as they would follow if extending the definition of the discrete symmetries 
from $d=(3+1)$ to any even $d$: (${\underline {\bf \mathbb{C}}}_{{{\bf \cal H}}}$,  ${\cal P}_{{\cal H}}$ 
and ${\cal T}_{{\cal H}}$), presented in Eqs.~(\ref{calCH}, \ref{calTPH}).  The discrete symmetries
(${\underline {\bf \mathbb{C}}}_{{{\bf \cal H}}}$,  ${\cal P}_{{\cal H}}$ and ${\cal T}_{{\cal H}}$) 
do not lead, namely, to the 
experimentally observed definitions,  since if using ${\underline {\bf \mathbb{C}}}_{{{\bf \cal H}}}$ 
on a second quantized 
state $\mathbf{\Psi}^{\dagger}$,  the charge conjugated state has the same charge as the starting state. 
The proposed new discrete symmetries  (${\underline {\bf \mathbb{C}}}_{{{\bf \cal N}}}$,  
${\cal P}_{{\cal N}}$  
and ${\cal T}_{{\cal N}}$) behave as they should - in agreement with the observed properties of fermions
and anti-fermions.

We do not study in this paper the break of 
${\underline {\bf \mathbb{C}}}_{{{\bf \cal N}}}$,  ${\cal P}_{{\cal N}}$ and ${\cal T}_{{\cal N}}$ 
symmetries.

We analyse properties of the proposed symmetries from the point of view of the 
observables in $d=(3+1)$.
Our definition of discrete symmetries is, as discussed in  
this paper and in particular in sect. \ref{understanding}, more general and valid for spaces with the 
central points and rotational symmetries around these points and might be helpful also for finding 
appropriate discrete symmetries operators in examples, when compactification is made by  a torus, 
where  the generators of translations around the torus are declared 
as charges in $(3+1)$. 

These discrete symmetries do not distinguish among families of fermions as long as the family groups  
form equivalent representations with respect to the charge groups.

We illustrate our definition of the discrete symmetries on two cases: i. $d=(5+1)$ 
and ii. $d=(13+1)$. 
The first case is a toy model on which we show~\cite{NH,hn0203,DHN} that the Kaluza-Klein kind of 
theories can lead 
in non-compact spaces to observable (almost massless) properties of fermions. We present in 
Table~\ref{Table I.} one family of fermions of positive and negative energy states. 
We also presented a way for a possible compactification in this toy model to demonstrate that 
our definition of the discrete symmetries is meaningful~\ref{comp5+1}.

For the second illustration of the proposed discrete symmetries the one family spinor representation of the 
{\it spin-charge-family} theory, which explains the assumptions of the {\it standard model}, is taken. 
We present in Table~\ref{Table II.} the representation of quarks of particular colour charge, in 
Table~\ref{Table so13+1.} we present all the members of one representation. It contains quarks and leptons and
the corresponding charge conjugated states. 

The discrete 
symmetries proceed similarly to the case of $d=(5+1)$. 
In this second illustration fermions carry the experimentally recognized properties: $\mathbb{C}_{{\cal N}}$  
${\cal P}_{{\cal N}}$ transforms the right handed 
$u$-quark with the spin up, weak chargeless and of the  colour charge ($\frac{1}{2}, \frac{1}{2 \sqrt{3}}$) 
and the hyper charge equal to $\frac{2}{3}$ into 
the left handed weak chargeless anti-quark with spin up and with the anti-colour charge 
($-\frac{1}{2}, -\frac{1}{2 \sqrt{3}}$)  and anti-hyper charge $-\frac{2}{3}$ 
(see Appendix~\ref{technique} lines 1 and 35 and also the ref.~\cite{Portoroz03}, Table 2. line 1 and 
Table 4. line 35). $\mathbb{C}_{{\cal N}}$  ${\cal P}_{{\cal N}}$ transforms the weak charged 
($\frac{1}{2}$) left handed neutrino, with spin up and colour chargeless into the right handed weak 
anti-charged ($-\frac{1}{2}$) anti-neutrino with the spin up, anti-colour chargeless (see 
Appendix~\ref{technique} Table~\ref{Table so13+1.}, line 31 and 61 and 
also the ref.~\cite{Portoroz03}, Table 3, line 31 and Table 5, line  61).

We also discuss about an acceptable compactification procedure, which leads in this case to
the {\it standard model} as a low energy effective theory of the {\it spin-charge-family} theory. 
This study is in progress.

We concentrated on discrete symmetries of fermions, but discussed also the properties of bosonic fields in 
higher dimensions, which are assumed to be treated as background fields, discussing in sect.~\ref{understanding}
their  behaving with respect to both kinds of the discrete symmetries: 
${\underline {\bf \mathbb{C}}}_{{\bf {\cal N}}}$, ${\cal P}^{(d-1)}_{{\cal N}}$  
and ${\cal T}_{{\cal N}}$ and ${\underline {\bf \mathbb{C}}}_{{\bf {\cal H}}}$, ${\cal P}^{(d-1)}_{{\cal H}}$   
and ${\cal T}_{{\cal H}}$.

The proposed discrete symmetries ${\underline {\bf \mathbb{C}}}_{{\bf {\cal N}}}$,  
${\cal P}^{(d-1)}_{{\cal N}}$ and ${\cal T}_{{\cal N}}$,
defined for spaces with dimensions $d$ even 
have obviously the desired properties in the observable part of space in cases with central point symmetries 
and the rotational symmetries around such central points~\cite{kk,zelenaknjiga,witten,sugra,norma,NPLB}, in which the 
way of curling up the higher dimensional space into  (almost) compact spaces or non compact spaces 
do not break a parity.

To discuss discrete symmetries of Kaluza-Klein kind of theories proposed in the 
literature~\cite{cptKK,BRG}  from the point of view of our proposal would require our complete 
understanding of these models and in addition discussions with the authors.

\appendix*

\section{The technique for representing spinors~\cite{norma93,hn0203,NPLB}, a shortened version 
of the one presented in~\cite{NPLB}}
\label{technique}

The technique~\cite{norma93,hn0203,NPLB} can be used to construct a spinor basis for any dimension $d$
and any signature in an easy and transparent way. Equipped with the graphic presentation of basic states,  
the technique offers an elegant way to see all the quantum numbers of states with respect to the  
Lorentz groups, as well as transformation properties of the states under any Clifford algebra object.

The objects $\gamma^a$ 
have properties 
%
$\{ \gamma^a, \gamma^b\}_{+} = 2\eta^{ab}\,I, $
%
%
for any $d$, even or odd.  $I$ is the unit element in the Clifford algebra.

The Clifford algebra objects $S^{ab}$  
close the algebra of the Lorentz group 
%
$S^{ab}:  = (i/4) (\gamma^a \gamma^b - \gamma^b \gamma^a)\,$, 
$\{S^{ab},S^{cd}\}_{-}  = i(\eta^{ad} S^{bc} + \eta^{bc} S^{ad} - \eta^{ac} S^{bd} - \eta^{bd} S^{ac})\,$.
%
The ``Hermiticity'' property for $\gamma^a$'s:   
$\,\gamma^{a\dagger} = \eta^{aa} \gamma^a\,$ 
is assumed  in order that $\gamma^a$ are 
%
formally unitary, 
i.e. $\gamma^{a \,\dagger} \,\gamma^a=I$. 
%

The Cartan subalgebra of the algebra 
is chosen in even dimensional spaces  as follows: 
%
$\,S^{03}, S^{12}, S^{56}, \cdots, S^{d-1\; d}, \quad {\rm if } \quad d  = 2n \ge 4$.
%

The choice for  the Cartan subalgebra in $d > 4$ is straightforward.
It is  useful  to define one of the Casimir operators of the Lorentz group -  
the  handedness $\Gamma$ ($\{\Gamma, S^{ab}\}_- =0$) in any $d$, for  even dimensional spaces  it follows: 
%
$\Gamma^{(d)} :=(i)^{d/2}\; \;\;\;\;\;\prod_a \quad (\sqrt{\eta^{aa}} \gamma^a), \quad {\rm if } \quad d = 2n\,$. 
%
The product of $\gamma^a$'s in the ascending order with respect to 
the index $a$: $\gamma^0 \gamma^1\cdots \gamma^d$ is understood. 
It follows 
for any choice of the signature $\eta^{aa}$ that
$\Gamma^{\dagger}= \Gamma,\;
\Gamma^2 = I.$
For $d$ even the handedness  anticommutes with the Clifford algebra objects 
$\gamma^a$ ($\{\gamma^a, \Gamma \}_+ = 0$).
%

To make the technique simple  the graphic presentation is introduced
\begin{eqnarray}
\stackrel{ab}{(k)}:&=& 
\frac{1}{2}(\gamma^a + \frac{\eta^{aa}}{ik} \gamma^b)\,,\quad \quad
\stackrel{ab}{[k]}:=
\frac{1}{2}(1+ \frac{i}{k} \gamma^a \gamma^b)\,,
\label{signature}
\end{eqnarray}
where $k^2 = \eta^{aa} \eta^{bb}$.
One can easily check by taking into account the  Clifford algebra relation 
and the definition of $S^{ab}$ 
that if one multiplies from the left hand side by $S^{ab}$ 
the Clifford algebra objects $\stackrel{ab}{(k)}$ and $\stackrel{ab}{[k]}$, it follows that
\begin{eqnarray}
        S^{ab}\, \stackrel{ab}{(k)}= \frac{1}{2}\,k\, \stackrel{ab}{(k)}\,,\quad \quad 
        S^{ab}\, \stackrel{ab}{[k]}= \frac{1}{2}\,k \,\stackrel{ab}{[k]}\,,
\label{grapheigen}
\end{eqnarray}
which means that we get the same objects back multiplied by the constant $\frac{1}{2}k$.
This also means that when 
$\stackrel{ab}{(k)}$ and $\stackrel{ab}{[k]}$ act from the left hand side on  a
vacuum state $|\psi_0\rangle$ the obtained states are the eigenvectors of $S^{ab}$.
One can further recognize 
that $\gamma^a$ transform  $\stackrel{ab}{(k)}$ into  $\stackrel{ab}{[-k]}$, never to $\stackrel{ab}{[k]}$:  
\begin{eqnarray} 
\gamma^a \stackrel{ab}{(k)}&=& \eta^{aa}\stackrel{ab}{[-k]},\; 
\gamma^b \stackrel{ab}{(k)}= -ik \stackrel{ab}{[-k]}, \; 
\gamma^a \stackrel{ab}{[k]}= \stackrel{ab}{(-k)},\; 
\gamma^b \stackrel{ab}{[k]}= -ik \eta^{aa} \stackrel{ab}{(-k)}\,.
\label{snmb:gammatildegamma}
\end{eqnarray}
%

Let us deduce some useful relations
\begin{eqnarray}
\stackrel{ab}{(k)}\stackrel{ab}{(k)}& =& 0\,, \quad \quad \stackrel{ab}{(k)}\stackrel{ab}{(-k)}
= \eta^{aa}  \stackrel{ab}{[k]}\,, \quad  
 \stackrel{ab}{[k]}\stackrel{ab}{[k]} = \stackrel{ab}{[k]}\,, \quad \quad
\stackrel{ab}{[k]}\stackrel{ab}{[-k]}= 0\,,
 \nonumber\\
 \stackrel{ab}{(k)}\stackrel{ab}{[k]}& =& 0\,,\quad \quad \quad \stackrel{ab}{[k]}\stackrel{ab}{(k)}
=  \stackrel{ab}{(k)}\,, \quad \quad 
 \stackrel{ab}{(k)}\stackrel{ab}{[-k]}=  \stackrel{ab}{(k)}\,,
\quad \quad \stackrel{ab}{[k]}\stackrel{ab}{(-k)} =0\,. 
\label{graphbinoms}
\end{eqnarray}
%

Taking into account the above equations it is easy to find a Weyl spinor irreducible representation
for $d$-dimensional space. 

For $d$ even we simply make a starting state as a product of $d/2$, let us say, only nilpotents 
$\stackrel{ab}{(k)}$, one for each $S^{ab}$ of the Cartan subalgebra  elements, 
applying it on an (unimportant) vacuum state. 
Then the generators $S^{ab}$, which do not belong 
to the Cartan subalgebra, being applied on the starting state from the left, 
 generate all the members of one
Weyl spinor.  
\begin{eqnarray}
\stackrel{0d}{(k_{0d})} \stackrel{12}{(k_{12})} \stackrel{35}{(k_{35})}\cdots \stackrel{d-1\;d-2}{(k_{d-1\;d-2})}
\psi_0 \nonumber\\
\stackrel{0d}{[-k_{0d}]} \stackrel{12}{[-k_{12}]} \stackrel{35}{(k_{35})}\cdots \stackrel{d-1\;d-2}{(k_{d-1\;d-2})}
\psi_0 \nonumber\\
%
\vdots \nonumber\\
\stackrel{od}{(k_{0d})} \stackrel{12}{[-k_{12}]} \stackrel{35}{[-k_{35}]}\cdots \stackrel{d-1\;d-2}{(k_{d-1\;d-2})}
\psi_0 \nonumber\\
\vdots 
\label{graphicd}
\end{eqnarray}
All the states have the handedness $\Gamma $, since $\{ \Gamma, S^{ab}\}_{-} = 0$. 
States, belonging to one multiplet  with respect to the group $SO(q,d-q)$, that is to one
irreducible representation of spinors (one Weyl spinor), can have any phase. We made a choice
of the simplest one, taking all  phases equal to one.


There are two kinds of the Clifford algebra objects~\cite{norma93,NJMP}: besides the Dirac $\gamma^a$  
ones also $\tilde{\gamma}^a$, 
with the properties~\cite{norma93} 
\begin{eqnarray}
\label{gammatildegamma}&& 
%
\{ \tilde{\gamma}^a, \tilde{\gamma}^b\}_{+}= 2\eta^{ab}\,, \quad \quad
\{ \gamma^a, \tilde{\gamma}^b\}_{+} = 0\,,
\end{eqnarray}
for any $d$, even or odd.   $\tilde{\gamma}^a$ form the equivalent representations with respect to 
$\gamma^a$.
If $\gamma^a$ multiply any Clifford algebra object $ {\bf B} = \sum_{i=0,d}\, a_{a_1 \cdots a_i}
\gamma^{a_1} \cdots \gamma^{a_i}$ from the left hand side ($\gamma^a {\bf B}|vac>_{fam}\,,|vac>_{fam} $ 
is the vacuum state),
multiply $\tilde{\gamma}^a$ the same ${\bf B}$ from the right hand side ($\tilde{\gamma}^a {\bf B}|vac>_{fam}=
i(-)^{n_{B}}\,{\bf B}\gamma^a|vac>_{fam} $. $(-)^{n_{B}}=+1,-1$, when the object ${\bf B}$ has an even or odd 
Clifford  character, respectively. 

Correspondingly transforms $\gamma^a$ the object $\stackrel{ab}{(k)}$ into $\stackrel{ab}{[-k]}$, never to 
$\stackrel{ab}{[k]}$, and $\stackrel{ab}{[k]}$ to $\stackrel{ab}{(-k)}$, never to $\stackrel{ab}{(k)}$,
while $\tilde{\gamma}^a$ transforms $\stackrel{ab}{(k)}$ into $\stackrel{ab}{[k]}$, never to 
$\stackrel{ab}{[-k]}$, and $\stackrel{ab}{[k]}$ to $\stackrel{ab}{(k)}$, never to $\stackrel{ab}{(-k)}$. 

The generators $S^{ab}$ (or superposition of $S^{ab}$) take care of spins and charges of the family members,
while $\tilde{S}^{ab}$ (or superposition of $\tilde{S}^{ab}$) take care of families.

We present below the multiplet of states and charge conjugated states of quarks and leptons belonging to 
one $SO(13,1)$ multiplet of quarks and leptons. To find families of the subgroup $SO(7,1)$ the generators 
of $\tilde{S}^{ab}, (a,b) \in (0,..,7,8)$ , with the property $\{\tilde{S}^{ab},S^{ab}\}_{-}=0 $ must be taken 
into account. 
%
\newpage
\bottomcaption{\label{Table so13+1.}
The left handed ($\Gamma^{(13,11)} = -1 = \Gamma^{(7,1)} 
\times \Gamma^{(6)}$) multiplet of spinors - the members of the $SO(13,1)$ group, 
manifesting the subgroup $SO(7,1)$ - of the colour charged quarks and anti-quarks and the colourless 
leptons and anti-leptons is presented in the massless basis using the technique of this Appendix. 
It contains the left handed  ($\Gamma^{(3,1)}$) weak charged  ($\tau^{13}$) 
and $SU(2)_{II}$ chargeless ($\tau^{23}$) 
quarks and the right handed weak chargeless and $SU(2)_{II}$ charged 
quarks of three colours  ($c^i$ $= (\tau^{33}, \tau^{38})$), 
with the spinor charge ($\tau^{4}$) 
and the colourless left handed weak charged leptons and the right handed weak chargeless leptons.  
$ S^{12}$ defines the ordinary spin 
$\pm \frac{1}{2}$. Additional notations are presented in Table~\ref{Table II.}.
The vacuum state $|vac-{fam}>$, on which the nilpotents and projectors operate, is not shown. 
The reader can find this  Weyl representation also in the refs.~\cite{Portoroz03,NJMP}. 
Two anti-octets of anti-quarks of the rest two anti-triplet colours follow from the presented one by substituting 
the part $||\stackrel{9 \;10}{[-]}\;\;\stackrel{11\;12}{(+)}\;\;\stackrel{13\;14}{(+)}$ of the basis by 
$||\stackrel{9 \;10}{(+)}\;\;\stackrel{11\;12}{[-]}\;\;\stackrel{13\;14}{(+)}$ for $\bar{q}_{L,R}^{\bar{c^2}}$
and by $||\stackrel{9 \;10}{(+)}\;\;\stackrel{11\;12}{(+)}\;\;\stackrel{13\;14}{[-]}$ for 
$\bar{q}_{L,R}^{\bar{c^3}}$. Correspondingly the charges are ($\frac{1}{2}, - \frac{1}{2 \sqrt{3}}$) and 
($0,  \frac{1}{ \sqrt{3}}$), respectively.}
%
\tablehead{\hline
i&$$&$|^a\psi_i>$&$\Gamma^{(3,1)}$&$ S^{12}$&$\Gamma^{(4)}$&
$\tau^{13}$&$\tau^{23}$&$\tau^{31}$&$\tau^{38}$&$\tau^{4}$&$Y$&$Q$\\
\hline
&& ${\rm Octet},\;\Gamma^{(1,7)} =1,\;\Gamma^{(6)} = -1,$&&&&&&& \\
&& ${\rm of \; quarks \;and \;leptons}$&&&&&&&\\
\hline\hline} 
\tabletail{\hline \multicolumn{13}{r}{\emph{Continued on next page}}\\}
\tablelasttail{\hline}
\begin{tiny}
\begin{center}
\begin{supertabular}{|r|c||c||c|c||c|c|c||c|c|c||r|r|}
1&$ u_{R}^{c1}$&$ \stackrel{03}{(+i)}\,\stackrel{12}{(+)}|
\stackrel{56}{(+)}\,\stackrel{78}{(+)}
||\stackrel{9 \;10}{(+)}\;\;\stackrel{11\;12}{[-]}\;\;\stackrel{13\;14}{[-]} $ &1&$\frac{1}{2}$&1&0&
$\frac{1}{2}$&$\frac{1}{2}$&$\frac{1}{2\,\sqrt{3}}$&$\frac{1}{6}$&$\frac{2}{3}$&$\frac{2}{3}$\\
\hline 
2&$u_{R}^{c1}$&$\stackrel{03}{[-i]}\,\stackrel{12}{[-]}|\stackrel{56}{(+)}\,\stackrel{78}{(+)}
||\stackrel{9 \;10}{(+)}\;\;\stackrel{11\;12}{[-]}\;\;\stackrel{13\;14}{[-]}$&1&$-\frac{1}{2}$&1&0&
$\frac{1}{2}$&$\frac{1}{2}$&$\frac{1}{2\,\sqrt{3}}$&$\frac{1}{6}$&$\frac{2}{3}$&$\frac{2}{3}$\\
\hline
3&$d_{R}^{c1}$&$\stackrel{03}{(+i)}\,\stackrel{12}{(+)}|\stackrel{56}{[-]}\,\stackrel{78}{[-]}
||\stackrel{9 \;10}{(+)}\;\;\stackrel{11\;12}{[-]}\;\;\stackrel{13\;14}{[-]}$&1&$\frac{1}{2}$&1&0&
$-\frac{1}{2}$&$\frac{1}{2}$&$\frac{1}{2\,\sqrt{3}}$&$\frac{1}{6}$&$-\frac{1}{3}$&$-\frac{1}{3}$\\
\hline 
4&$ d_{R}^{c1} $&$\stackrel{03}{[-i]}\,\stackrel{12}{[-]}|
\stackrel{56}{[-]}\,\stackrel{78}{[-]}
||\stackrel{9 \;10}{(+)}\;\;\stackrel{11\;12}{[-]}\;\;\stackrel{13\;14}{[-]} $&1&$-\frac{1}{2}$&1&0&
$-\frac{1}{2}$&$\frac{1}{2}$&$\frac{1}{2\,\sqrt{3}}$&$\frac{1}{6}$&$-\frac{1}{3}$&$-\frac{1}{3}$\\
\hline
5&$d_{L}^{c1}$&$\stackrel{03}{[-i]}\,\stackrel{12}{(+)}|\stackrel{56}{[-]}\,\stackrel{78}{(+)}
||\stackrel{9 \;10}{(+)}\;\;\stackrel{11\;12}{[-]}\;\;\stackrel{13\;14}{[-]}$&-1&$\frac{1}{2}$&-1&
$-\frac{1}{2}$&0&$\frac{1}{2}$&$\frac{1}{2\,\sqrt{3}}$&$\frac{1}{6}$&$\frac{1}{6}$&$-\frac{1}{3}$\\
\hline
6&$d_{L}^{c1} $&$\stackrel{03}{(+i)}\,\stackrel{12}{[-]}|\stackrel{56}{[-]}\,\stackrel{78}{(+)}
||\stackrel{9 \;10}{(+)}\;\;\stackrel{11\;12}{[-]}\;\;\stackrel{13\;14}{[-]} $&-1&$-\frac{1}{2}$&-1&
$-\frac{1}{2}$&0&$\frac{1}{2}$&$\frac{1}{2\,\sqrt{3}}$&$\frac{1}{6}$&$\frac{1}{6}$&$-\frac{1}{3}$\\
\hline
7&$ u_{L}^{c1}$&$\stackrel{03}{[-i]}\,\stackrel{12}{(+)}|\stackrel{56}{(+)}\,\stackrel{78}{[-]}
||\stackrel{9 \;10}{(+)}\;\;\stackrel{11\;12}{[-]}\;\;\stackrel{13\;14}{[-]}$ &-1&$\frac{1}{2}$&-1&
$\frac{1}{2}$&0 &$\frac{1}{2}$&$\frac{1}{2\,\sqrt{3}}$&$\frac{1}{6}$&$\frac{1}{6}$&$\frac{2}{3}$\\
\hline
8&$u_{L}^{c1}$&$\stackrel{03}{(+i)}\,\stackrel{12}{[-]}|\stackrel{56}{(+)}\,\stackrel{78}{[-]}
||\stackrel{9 \;10}{(+)}\;\;\stackrel{11\;12}{[-]}\;\;\stackrel{13\;14}{[-]}$&-1&$-\frac{1}{2}$&-1&
$\frac{1}{2}$&0&$\frac{1}{2}$&$\frac{1}{2\,\sqrt{3}}$&$\frac{1}{6}$&$\frac{1}{6}$&$\frac{2}{3}$\\
\hline\hline
9&$ u_{R}^{c2}$&$ \stackrel{03}{(+i)}\,\stackrel{12}{(+)}|
\stackrel{56}{(+)}\,\stackrel{78}{(+)}
||\stackrel{9 \;10}{[-]}\;\;\stackrel{11\;12}{(+)}\;\;\stackrel{13\;14}{[-]} $ &1&$\frac{1}{2}$&1&0&
$\frac{1}{2}$&$-\frac{1}{2}$&$\frac{1}{2\,\sqrt{3}}$&$\frac{1}{6}$&$\frac{2}{3}$&$\frac{2}{3}$\\
\hline 
10&$u_{R}^{c2}$&$\stackrel{03}{[-i]}\,\stackrel{12}{[-]}|\stackrel{56}{(+)}\,\stackrel{78}{(+)}
||\stackrel{9 \;10}{[-]}\;\;\stackrel{11\;12}{(+)}\;\;\stackrel{13\;14}{[-]}$&1&$-\frac{1}{2}$&1&0&
$\frac{1}{2}$&$-\frac{1}{2}$&$\frac{1}{2\,\sqrt{3}}$&$\frac{1}{6}$&$\frac{2}{3}$&$\frac{2}{3}$\\
\hline
$\cdots$&&&&&&&&&&&&\\
\hline
\hline\hline
17&$ u_{R}^{c3}$&$ \stackrel{03}{(+i)}\,\stackrel{12}{(+)}|
\stackrel{56}{(+)}\,\stackrel{78}{(+)}
||\stackrel{9 \;10}{[-]}\;\;\stackrel{11\;12}{[-]}\;\;\stackrel{13\;14}{(+)} $ &1&$\frac{1}{2}$&1&0&
$\frac{1}{2}$&$0$&$-\frac{1}{\sqrt{3}}$&$\frac{1}{6}$&$\frac{2}{3}$&$\frac{2}{3}$\\
\hline 
18&$u_{R}^{c3}$&$\stackrel{03}{[-i]}\,\stackrel{12}{[-]}|\stackrel{56}{(+)}\,\stackrel{78}{(+)}
||\stackrel{9 \;10}{[-]}\;\;\stackrel{11\;12}{[-]}\;\;\stackrel{13\;14}{(+)}$&1&$-\frac{1}{2}$&1&0&
$\frac{1}{2}$&$0$&$-\frac{1}{\sqrt{3}}$&$\frac{1}{6}$&$\frac{2}{3}$&$\frac{2}{3}$\\
\hline
$\cdots$&&&&&&&&&&&&\\
\hline\hline
25&$ \nu_{R}$&$ \stackrel{03}{(+i)}\,\stackrel{12}{(+)}|
\stackrel{56}{(+)}\,\stackrel{78}{(+)}
||\stackrel{9 \;10}{(+)}\;\;\stackrel{11\;12}{(+)}\;\;\stackrel{13\;14}{(+)} $ &1&$\frac{1}{2}$&1&0&
$\frac{1}{2}$&$0$&$0$&$-\frac{1}{2}$&$0$&$0$\\
\hline 
26&$\nu_{R}$&$\stackrel{03}{[-i]}\,\stackrel{12}{[-]}|\stackrel{56}{(+)}\,\stackrel{78}{(+)}
||\stackrel{9 \;10}{(+)}\;\;\stackrel{11\;12}{(+)}\;\;\stackrel{13\;14}{(+)}$&1&$-\frac{1}{2}$&1&0&
$\frac{1}{2}$ &$0$&$0$&$-\frac{1}{2}$&$0$&$0$\\
\hline
27&$e_{R}$&$\stackrel{03}{(+i)}\,\stackrel{12}{(+)}|\stackrel{56}{[-]}\,\stackrel{78}{[-]}
||\stackrel{9 \;10}{(+)}\;\;\stackrel{11\;12}{(+)}\;\;\stackrel{13\;14}{(+)}$&1&$\frac{1}{2}$&1&0&
$-\frac{1}{2}$&$0$&$0$&$-\frac{1}{2}$&$-1$&$-1$\\
\hline 
28&$ e_{R} $&$\stackrel{03}{[-i]}\,\stackrel{12}{[-]}|
\stackrel{56}{[-]}\,\stackrel{78}{[-]}
||\stackrel{9 \;10}{(+)}\;\;\stackrel{11\;12}{(+)}\;\;\stackrel{13\;14}{(+)} $&1&$-\frac{1}{2}$&1&0&
$-\frac{1}{2}$&$0$&$0$&$-\frac{1}{2}$&$-1$&$-1$\\
\hline
29&$e_{L}$&$\stackrel{03}{[-i]}\,\stackrel{12}{(+)}|\stackrel{56}{[-]}\,\stackrel{78}{(+)}
||\stackrel{9 \;10}{(+)}\;\;\stackrel{11\;12}{(+)}\;\;\stackrel{13\;14}{(+)}$&-1&$\frac{1}{2}$&-1&
$-\frac{1}{2}$&0&$0$&$0$&$-\frac{1}{2}$&$-\frac{1}{2}$&$-1$\\
\hline
30&$e_{L} $&$\stackrel{03}{(+i)}\,\stackrel{12}{[-]}|\stackrel{56}{[-]}\,\stackrel{78}{(+)}
||\stackrel{9 \;10}{(+)}\;\;\stackrel{11\;12}{(+)}\;\;\stackrel{13\;14}{(+)} $&-1&$-\frac{1}{2}$&-1&
$-\frac{1}{2}$&0&$0$&$0$&$-\frac{1}{2}$&$-\frac{1}{2}$&$-1$\\
\hline
31&$ \nu_{L}$&$\stackrel{03}{[-i]}\,\stackrel{12}{(+)}|\stackrel{56}{(+)}\,\stackrel{78}{[-]}
||\stackrel{9 \;10}{(+)}\;\;\stackrel{11\;12}{(+)}\;\;\stackrel{13\;14}{(+)}$ &-1&$\frac{1}{2}$&-1&
$\frac{1}{2}$&0 &$0$&$0$&$-\frac{1}{2}$&$-\frac{1}{2}$&$0$\\
\hline
32&$\nu_{L}$&$\stackrel{03}{(+i)}\,\stackrel{12}{[-]}|\stackrel{56}{(+)}\,\stackrel{78}{[-]}
||\stackrel{9 \;10}{(+)}\;\;\stackrel{11\;12}{(+)}\;\;\stackrel{13\;14}{(+)}$&-1&$-\frac{1}{2}$&-1&
$\frac{1}{2}$&0&$0$&$0$&$-\frac{1}{2}$&$-\frac{1}{2}$&$0$\\
\hline\hline
33&$ \bar{d}_{L}^{\bar{c1}}$&$ \stackrel{03}{[-i]}\,\stackrel{12}{(+)}|
\stackrel{56}{(+)}\,\stackrel{78}{(+)}
||\stackrel{9 \;10}{[-]}\;\;\stackrel{11\;12}{(+)}\;\;\stackrel{13\;14}{(+)} $ &-1&$\frac{1}{2}$&1&0&
$\frac{1}{2}$&$-\frac{1}{2}$&$-\frac{1}{2\,\sqrt{3}}$&$-\frac{1}{6}$&$\frac{1}{3}$&$\frac{1}{3}$\\
\hline 
34&$\bar{d}_{L}^{\bar{c1}}$&$\stackrel{03}{(+i)}\,\stackrel{12}{[-]}|\stackrel{56}{(+)}\,\stackrel{78}{(+)}
||\stackrel{9 \;10}{[-]}\;\;\stackrel{11\;12}{(+)}\;\;\stackrel{13\;14}{(+)}$&-1&$-\frac{1}{2}$&1&0&
$\frac{1}{2}$&$-\frac{1}{2}$&$-\frac{1}{2\,\sqrt{3}}$&$-\frac{1}{6}$&$\frac{1}{3}$&$\frac{1}{3}$\\
\hline
35&$\bar{u}_{L}^{\bar{c1}}$&$\stackrel{03}{[-i]}\,\stackrel{12}{(+)}|\stackrel{56}{[-]}\,\stackrel{78}{[-]}
||\stackrel{9 \;10}{[-]}\;\;\stackrel{11\;12}{(+)}\;\;\stackrel{13\;14}{(+)}$-&1&$\frac{1}{2}$&1&0&
$-\frac{1}{2}$&$-\frac{1}{2}$&$-\frac{1}{2\,\sqrt{3}}$&$-\frac{1}{6}$&$-\frac{2}{3}$&$-\frac{2}{3}$\\
\hline
36&$ \bar{u}_{L}^{\bar{c1}} $&$\stackrel{03}{(+i)}\,\stackrel{12}{[-]}|
\stackrel{56}{[-]}\,\stackrel{78}{[-]}
||\stackrel{9 \;10}{[-]}\;\;\stackrel{11\;12}{(+)}\;\;\stackrel{13\;14}{(+)} $&-1&$-\frac{1}{2}$&1&0&
$-\frac{1}{2}$&$-\frac{1}{2}$&$-\frac{1}{2\,\sqrt{3}}$&$-\frac{1}{6}$&$-\frac{2}{3}$&$-\frac{2}{3}$\\
\hline
37&$\bar{d}_{R}^{\bar{c1}}$&$\stackrel{03}{(+i)}\,\stackrel{12}{(+)}|\stackrel{56}{(+)}\,\stackrel{78}{[-]}
||\stackrel{9 \;10}{[-]}\;\;\stackrel{11\;12}{(+)}\;\;\stackrel{13\;14}{(+)}$&1&$\frac{1}{2}$&-1&
$\frac{1}{2}$&0&$-\frac{1}{2}$&$-\frac{1}{2\,\sqrt{3}}$&$-\frac{1}{6}$&$-\frac{1}{6}$&$\frac{1}{3}$\\
\hline
38&$\bar{d}_{R}^{\bar{c1}} $&$\stackrel{03}{[-i]}\,\stackrel{12}{[-]}|\stackrel{56}{(+)}\,\stackrel{78}{[-]}
||\stackrel{9 \;10}{[-]}\;\;\stackrel{11\;12}{(+)}\;\;\stackrel{13\;14}{(+)} $&1&$-\frac{1}{2}$&-1&
$\frac{1}{2}$&0&$-\frac{1}{2}$&$-\frac{1}{2\,\sqrt{3}}$&$-\frac{1}{6}$&$-\frac{1}{6}$&$\frac{1}{3}$\\
\hline
39&$ \bar{u}_{R}^{\bar{c1}}$&$\stackrel{03}{(+i)}\,\stackrel{12}{(+)}|\stackrel{56}{[-]}\,\stackrel{78}{(+)}
||\stackrel{9 \;10}{[-]}\;\;\stackrel{11\;12}{(+)}\;\;\stackrel{13\;14}{(+)}$ &1&$\frac{1}{2}$&-1&
$-\frac{1}{2}$&0 &$-\frac{1}{2}$&$-\frac{1}{2\,\sqrt{3}}$&$-\frac{1}{6}$&$-\frac{1}{6}$&$-\frac{2}{3}$\\
\hline
40&$\bar{u}_{R}^{\bar{c1}}$&$\stackrel{03}{[-i]}\,\stackrel{12}{[-]}|\stackrel{56}{[-]}\,\stackrel{78}{(+)}
||\stackrel{9 \;10}{[-]}\;\;\stackrel{11\;12}{(+)}\;\;\stackrel{13\;14}{(+)}$&1&$-\frac{1}{2}$&-1&
$-\frac{1}{2}$&0&$-\frac{1}{2}$&$-\frac{1}{2\,\sqrt{3}}$&$-\frac{1}{6}$&$-\frac{1}{6}$&$-\frac{2}{3}$\\
\hline\hline
41&$ \bar{d}_{L}^{\bar{c2}}$&$ \stackrel{03}{[-i]}\,\stackrel{12}{(+)}|
\stackrel{56}{(+)}\,\stackrel{78}{(+)}
||\stackrel{9 \;10}{(+)}\;\;\stackrel{11\;12}{[-]}\;\;\stackrel{13\;14}{(+)} $ &-1&$\frac{1}{2}$&1&0&
$\frac{1}{2}$&$\frac{1}{2}$&$-\frac{1}{2\,\sqrt{3}}$&$-\frac{1}{6}$&$\frac{1}{3}$&$\frac{1}{3}$\\
\hline 
$\cdots$ &&&&&&&&&&&& \\
\hline\hline
49&$ \bar{d}_{L}^{\bar{c3}}$&$ \stackrel{03}{[-i]}\,\stackrel{12}{(+)}|
\stackrel{56}{(+)}\,\stackrel{78}{(+)}
||\stackrel{9 \;10}{(+)}\;\;\stackrel{11\;12}{(+)}\;\;\stackrel{13\;14}{[-]} $ &-1&$\frac{1}{2}$&1&0&
$\frac{1}{2}$&$0$&$-\frac{1}{\sqrt{3}}$&$-\frac{1}{6}$&$\frac{1}{3}$&$\frac{1}{3}$\\
\hline 
$\cdots$ &&&&&&&&&&&& \\
\hline\hline
57&$ \bar{e}_{L}$&$ \stackrel{03}{[-i]}\,\stackrel{12}{(+)}|
\stackrel{56}{(+)}\,\stackrel{78}{(+)}
||\stackrel{9 \;10}{[-]}\;\;\stackrel{11\;12}{[-]}\;\;\stackrel{13\;14}{[-]} $ &-1&$\frac{1}{2}$&1&0&
$\frac{1}{2}$&$0$&$0$&$\frac{1}{2}$&$1$&$1$\\
\hline 
58&$\bar{e}_{L}$&$\stackrel{03}{(+i)}\,\stackrel{12}{[-]}|\stackrel{56}{(+)}\,\stackrel{78}{(+)}
||\stackrel{9 \;10}{[-]}\;\;\stackrel{11\;12}{[-]}\;\;\stackrel{13\;14}{[-]}$&-1&$-\frac{1}{2}$&1&0&
$\frac{1}{2}$ &$0$&$0$&$\frac{1}{2}$&$1$&$1$\\
\hline
59&$\bar{\nu}_{L}$&$\stackrel{03}{[-i]}\,\stackrel{12}{(+)}|\stackrel{56}{[-]}\,\stackrel{78}{[-]}
||\stackrel{9 \;10}{[-]}\;\;\stackrel{11\;12}{[-]}\;\;\stackrel{13\;14}{[-]}$&-1&$\frac{1}{2}$&1&0&
$-\frac{1}{2}$&$0$&$0$&$\frac{1}{2}$&$0$&$0$\\
\hline 
60&$ \bar{\nu}_{L} $&$\stackrel{03}{(+i)}\,\stackrel{12}{[-]}|
\stackrel{56}{[-]}\,\stackrel{78}{[-]}
||\stackrel{9 \;10}{[-]}\;\;\stackrel{11\;12}{[-]}\;\;\stackrel{13\;14}{[-]} $&-1&$-\frac{1}{2}$&1&0&
$-\frac{1}{2}$&$0$&$0$&$\frac{1}{2}$&$0$&$0$\\
\hline
61&$\bar{\nu}_{R}$&$\stackrel{03}{(+i)}\,\stackrel{12}{(+)}|\stackrel{56}{[-]}\,\stackrel{78}{(+)}
||\stackrel{9 \;10}{[-]}\;\;\stackrel{11\;12}{[-]}\;\;\stackrel{13\;14}{[-]}$&1&$\frac{1}{2}$&-1&
$-\frac{1}{2}$&0&$0$&$0$&$\frac{1}{2}$&$\frac{1}{2}$&$0$\\
\hline
62&$\bar{\nu}_{R} $&$\stackrel{03}{[-i]}\,\stackrel{12}{[-]}|\stackrel{56}{[-]}\,\stackrel{78}{(+)}
||\stackrel{9 \;10}{[-]}\;\;\stackrel{11\;12}{[-]}\;\;\stackrel{13\;14}{[-]} $&1&$-\frac{1}{2}$&-1&
$-\frac{1}{2}$&0&$0$&$0$&$\frac{1}{2}$&$\frac{1}{2}$&$0$\\
\hline
63&$ \bar{e}_{R}$&$\stackrel{03}{(+i)}\,\stackrel{12}{(+)}|\stackrel{56}{(+)}\,\stackrel{78}{[-]}
||\stackrel{9 \;10}{[-]}\;\;\stackrel{11\;12}{[-]}\;\;\stackrel{13\;14}{[-]}$ &1&$\frac{1}{2}$&-1&
$\frac{1}{2}$&0 &$0$&$0$&$\frac{1}{2}$&$\frac{1}{2}$&$1$\\
\hline
64&$\bar{e}_{R}$&$\stackrel{03}{[-i]}\,\stackrel{12}{[-]}|\stackrel{56}{(+)}\,\stackrel{78}{[-]}
||\stackrel{9 \;10}{[-]}\;\;\stackrel{11\;12}{[-]}\;\;\stackrel{13\;14}{[-]}$&1&$-\frac{1}{2}$&-1&
$\frac{1}{2}$&0&$0$&$0$&$\frac{1}{2}$&$\frac{1}{2}$&$1$\\
\hline 
\end{supertabular}
\end{center}
\end{tiny}

\section*{Acknowledgments} The authors acknowledge funding of the Slovenian Research Agency and of the
Niels Bohr Institute (as a part of the emeritus status of H.B.N.) and in particular the financial support 
of the enterprise Beyond Semiconductors/BS Storitve d.o.o., Matja\v z Breskvar, to the Bled workshops
"What comes beyond the standard models", where this work has started.

\end{document}